\documentclass[12pt]{article}
\usepackage{amsmath}
\usepackage{graphicx,psfrag,epsf}
\usepackage{enumerate}
\usepackage{psfrag,epsf}
\usepackage{url} % not crucial - just used below for the URL
\usepackage{amsmath,amssymb,amsfonts,amsthm,mathtools,mathrsfs,float}
\usepackage{adjustbox}
\usepackage[sort, numbers]{natbib}
\usepackage{bm,bbm}
\usepackage{color}
\usepackage{subfigure}
\usepackage{algorithm,algpseudocode}
\usepackage[symbol]{footmisc}
\usepackage{scalefnt}
\usepackage{authblk} % author and affiliations
\usepackage{multirow,centernot}
\usepackage[colorlinks=true, citecolor=blue, urlcolor=blue]{hyperref}
\usepackage{natbib}
\usepackage{dsfont}
\usepackage[title]{appendix}
\input cyracc.def

\usepackage[left=1in,top=1.1in,right=0.5in,bottom=1in]{geometry}

\theoremstyle{definition}

\newtheorem{theorem}{Theorem}[section]

\newtheorem{remark}[theorem]{Remark}
\makeatletter
\def\@seccntformat#1{\@ifundefined{#1@cntformat}%
	{\csname the#1\endcsname\quad}%      default
	{\csname #1@cntformat\endcsname}%    enable individual control
}
\makeatother

\newif\ifShowComments
\ShowCommentstrue
\def\strutdepth{\dp\strutbox}
\def\druk#1{\strut\vadjust{\kern-\strutdepth
		{\vtop to \strutdepth{%
				\baselineskip\strutdepth\vss
				\llap{\hbox{#1}\quad}\null}}}}

\title{\bf
A new unit-bimodal distribution based on correlated Birnbaum-Saunders random variables
}

\author[1]{Roberto Vila \thanks{rovig161@gmail.com}}
\author[1]{Helton Saulo  \thanks{heltonsaulo@gmail.com} }
\author[1]{Felipe Quintino  \thanks{felipesquintino2@gmail.com} }
\author[1]{Peter  Zörnig  \thanks{peter@unb.br} }
\affil[1]{Department of Statistics, University of
	Bras\'ilia, 70910-900, Bras\'ilia, Brazil}
\begin{document}
	\maketitle 	
	\begin{abstract}
		{
In this paper, we propose a new distribution over the unit interval which can be characterized as a ratio of the type $Z\stackrel{d}{=}Y/(X+Y)$ where $X$ and $Y$ are two correlated Birnbaum-Saunders random variables. {\color{black} The density of $Z$ may be unimodal or bimodal. Simple expressions for the cumulative distribution function, moment-generating function and moments are obtained. Moreover,} the stress-strength probability between $X$ and $Y$ is calculated explicitly {\color{black} in the symmetric case, that is,} when the respective scale parameters are equal. Two applications of the ratio distribution are discussed.
		}

	\end{abstract}
	
	%{
		%	\hypersetup{linkcolor=black}
		%	\tableofcontents
		%}
	
\section{Introduction}
%\label{sec:1}
\noindent

Traditionally, the activities in statistical distribution theory have been concentrated on distributions with unbounded support. Only in recent years, attention has been devoted to the construction of distributions with bounded support. Bounded data have applications in economics, finance, biology and medicine among others, see e.g., the introductory remarks in  \cite{Vila2024a} and \cite{Vila2024b}. Such data arise naturally in the form of rates and proportions. Some of the most prominent bounded distributions are the beta distribution and their generalizations and the distributions of Kumaraswamy and Topp-Leone. 

A natural way to create a model for bounded data is to transform one or two unbounded random variables in a bounded variable $Z$. For example, $Z=X/(X+1)$, $Z=\exp(-X)$, {\color{black} $Z=\tanh(X)$, $Z=\min\{X,Y\}/\max\{X,Y\}$, $Z=(X/Y)\mathds{1}_{\{X<Y\}}$} and $Z=X/(X+Y)$ have the support $(0, 1)$, when $X$ and $Y$ are positive variables. In applications of the last model it is usually assumed that $X$ and $Y$ are independent. We consider some specific approaches using these transformations or more complicated ones. 
{\color{black} These techniques are simple transformations of one or two variables:}
\cite{Vila2024a} derive a bimodal beta distribution, applying a quadratic transformation technique due to Elal-Olivero to the beta distribution.
\cite{Afifi2022} apply an approach of Marshall and Olkin to the reduced Kies distribution.
\cite{Condino2023} propose a general framework in which a monotone function is applied to a single variable. 
\cite{Zornig2023} applies a procedure due to  \cite{Cortes2018} to a uniformly distributed variable, resulting in a three-parameter family of distributions over the unit interval.
The authors in \cite{Vila2024c} (unit log) use the approach $Z=X/(X+Y)$ where $X$ and $Y$ are the components of a two-dimensional random vector, following a bivariate log-symmetric distribution. In this model, $X$ and $Y$ may be correlated.

The distribution of ratios of the latter type is of great importance in practice, for example in modeling COVID-19-related death rates, see \cite{Bourguignon2024}.

On the other hand, the Birnbaum-Saunders (BS) fatigue-life model, introduced by \cite{Birnbaum1969a}, is based on renewal theory principles. This model focuses on the number of cycles needed to exceed a critical threshold and cause fatigue crack propagation. In the same year, \cite{Birnbaum1969b} provided maximum likelihood estimates for the BS model parameters. Later, Desmond (1985) \cite{Desmond1985}, in his work on stochastic models of failure in random environments, offered an alternative derivation of the distribution using a biological model and relaxed several of the assumptions made by \cite{Birnbaum1969a}. The BS distribution has been extensively studied and applied across various fields, including business, industry, economics, finance, management, engineering, and medical sciences. For a review of recent work on the BS distribution, see \cite{Saulo2019} and the references therein.
Bivariate (and multivariate) versions of the BS distribution have also been explored. \cite{Kundu2010} introduced the bivariate BS distribution with five parameters using the bivariate normal distribution function. \cite{Kundu2017} revisited the bivariate BS distribution and discussed several new properties. \cite{Saulo2019} introduced a mean-based bivariate BS distribution, demonstrating that the mean is one of its parameters.

In the present paper we propose the distribution of a random variable of the type $Z=Y/(X+Y)$, where $X$ and $Y$ are two (possibly correlated) Birnbaum-Saunders random variables. The article is structured as follows. 
In Section \ref{sec:1}, we introduce the unit{\color{black}-bimodal} Birnbaum-Saunders model{\color{black}, which we name by type 1 and denote by UBBS1 to differentiate from the unit-bimodal Birnbaum-Saunders (UBBS) distribution studied in \cite{Martinez2024unit},} and some of its forms are outlined in graphs. Note that bimodality in the model is observed as a function of its shape and correlation parameters. {\color{black} For a unit Birnbaum-Saunders model based on the transformation $Z=\exp(-X)$, with $X$ having a BS distribution,  see \cite{Mazucheli2018,Mazucheli2021}.
	Since the variable $Z$ in \cite{Mazucheli2018,Mazucheli2021} exhibits decreasing, upside down bathtub and bathtub shaped density,
	this model is not suitable to capture bimodal pattern in data.}
In Section \ref{sec:2}, {\color{black} we show one of the main properties of the UBBS1 model, that is, the corresponding random variable originally comes from} a type I ratio \citep{Johnson95,Bekker2009} of two variables following a bivariate BS distribution. 

Also in this section, we derive the cumulative distribution function (CDF), which is used to {\color{black} obtain simple expressions of the moment-generating function and moments, and to} explicitly calculate the stress-strength probability in the case of the presence of correlation and equality of scale parameters. {\color{black}Finally, the maximum likelihood and maximum product of spacings methods for parameter estimation are presented. In Section \ref{sec:04}, we carry out a Monte Carlo simulation study to evaluate the performance of the maximum likelihood and maximum product of spacings estimators by means of their bias and root mean square error. In this section, we also compare the performance of the estimation methods when bimodality is present.} In Section \ref{sec:05}, we present two applications to income-consumption and body mass data. %We showed that these data can also be fitted well by the {\color{black} UBBS1} distribution.
{\color{black} The first data set presents unimodal characteristics, and in this case, the new UBBS1 model proved to be more appropriate than the Beta and UBBS models (from \cite{Martinez2024unit}). The second data set has bimodal characteristics and was previously modeled by \cite{Martinez2024unit} using the UBBS model. Now, we show that the UBBS1 model also provides a good fit for this data.}
Finally, in Section \ref{sec:06}, some concluding remarks are made.

\section{{\color{black} The unit-bimodal Birnbaum-Saunders distribution of type 1}}
\label{sec:1}

We say that a continuous random variable $Z$, with support $(0, 1)$, follows a unit{\color{black}-bimodal} BS distribution {\color{black} of type 1} {\color{black} (UBBS1)} with parameter vector $\boldsymbol{\theta}_\rho=(\alpha_1,\alpha_2,\beta_1,\beta_2,\rho)^\top$, $\alpha_1> 0,\alpha_2> 0 ,\beta_1> 0, \beta_2> 0$ and $\rho\in(-1,1)$,
denoted by $Z\sim {\rm {\color{black} UBBS1}}(\boldsymbol{\theta}_\rho)$, if its probability density function (PDF) is given by (for $0<z<1$ and $s={1/z}-1$)
\begin{align}\label{main-pdf}
	\begin{array}{lll}
		f_Z(z;\boldsymbol{\theta}_\rho)
		&=
		{\exp\left\{{1\over 1-\rho^2}
			\left(
			{1\over\alpha_1^2}+{1\over\alpha_2^2}
			\right)\right\}\over 4\pi\alpha_1\alpha_2 \sqrt{1-\rho^2}}\, 
		{(s+1)^2\over s}\,
		\exp\left\{
		-
		{\rho\over \alpha_1\alpha_2(1-\rho^2)}
		\left(\sqrt{\beta_2 s\over\beta_1}
		+
		\sqrt{\beta_1\over \beta_2 s}\right)
		\right\}
		%\nonumber
		\\[0,5cm]
		&\times 
		\Bigg\{\!
		%\sqrt{\beta_1\beta_2\over sa_sb_s}\,
		%K_{-1}\left({1\over 1-\rho^2}\, \sqrt{{b_s\over a_s }}\right)
		%+
		\left(
		\sqrt{\beta_2 s\over\beta_1}
		+
		\sqrt{\beta_1\over \beta_2 s}
		\right) 
		K_{0}\left({\sqrt{{u_{\rho} v_{\rho} }}\over 1-\rho^2}\right)
		+
		\left[
		\sqrt{(v_{\rho}/u_{\rho})s\over\beta_1\beta_2}
		+
		\sqrt{\beta_1\beta_2\over (v_{\rho}/u_{\rho})s}\,
		\right] 
		K_{1}\left({\sqrt{{u_{\rho} v_{\rho} }}\over 1-\rho^2}
		\right)\!
		\Bigg\},
	\end{array}
\end{align}
where 
\begin{align}\label{def-a-b}
	u_{\rho}
	\equiv
	{s\over\alpha_1^2\beta_1}
	+
	{1\over\alpha_2^2\beta_2}
	-
	{2\rho\over\alpha_1\alpha_2}
	\sqrt{s\over\beta_1\beta_2},
	\quad 
	v_{\rho}
	\equiv 
	{\beta_1\over \alpha_1^2 s}
	+
	{\beta_2\over \alpha_2^2}
	-
	{2\rho\over\alpha_1\alpha_2}
	\sqrt{\beta_1\beta_2\over s}.
\end{align}
Furthermore, in formula \eqref{main-pdf},
\begin{align}\label{Bessel}
	K_\lambda(x)
	&=
	{1\over 2}\int_0^\infty w^{\lambda-1}
	\exp\left\{-{1\over 2}\, x\left(w+{1\over w}\right)\right\}
	{\rm d}w, \quad x>0, \lambda\in\mathbb{R},
	\nonumber
	\\[0,2cm]
	&=
	{\pi\over 2\sin(\lambda\pi)}\left[I_{-\lambda}(x)-I_{\lambda}(x)\right]
\end{align}
is the modified Bessel function of the third kind (also known as modified Bessel function of the second kind) \citep{Jorgensen1982,Abramowitz1972},
with
\begin{align*}
	I_{\lambda}(x)
	=
	\left({x\over 2}\right)^\lambda \sum_{k=0}^\infty {1\over k!\Gamma(\lambda+k+1)}\left({x\over 2}\right)^{2k}
\end{align*}
being the modified Bessel function of the first kind. When $\lambda\in\mathbb{Z}$ the following integral representation of $I_{\lambda}$ is valid \citep[][p. 376]{Abramowitz1972}:
$
I_{\lambda}(x)
=
({1/\pi}) \int_0^\pi \exp\{x\cos(\theta)\}\cos(\lambda\theta){\rm d}\theta.
$
{\color{black}
	\begin{remark}\label{rem-1}
		A simple algebraic manipulation shows that, when $\beta_1=\beta_2$, 	$f_Z(1/2 - w;\boldsymbol{\theta}_\rho)=f_Z(1/2 + w;\boldsymbol{\theta}_\rho)$, for all $0 < w < 1/2$. That is, The UBBS1 PDF in \eqref{main-pdf} is symmetric around $z_0 =1/2$ provided $\beta_1=\beta_2$ (see Figure \ref{fig:pdf_same_beta}).  Furthermore, in this case, the median and the mean of a UBBS1 distribution both occur at $z_0 = 1/2$.
	\end{remark}
}

\smallskip
When $\rho=0$, that is, $X$ and $Y$ are independent, the PDF \eqref{main-pdf} of $Z$ is simplified as (for $0<z<1$ and $s={1/z}-1$)
\begin{align}
	\begin{array}{lll}
		f_Z(z;\boldsymbol{\theta}_0)
		&=
		{\exp
			\left(
			{1\over\alpha_1^2}+{1\over\alpha_2^2}
			\right)\over 4\pi\alpha_1\alpha_2}\, 
		{(s+1)^2\over s}
		\\[0,5cm]
		&\times\Bigg\{
		\left(
		\sqrt{\beta_2 s\over\beta_1}
		+
		\sqrt{\beta_1\over \beta_2 s}
		\right) 
		K_{0}\left(\sqrt{{uv}}\right)
		+
		\left[
		\sqrt{(v/u)s\over\beta_1\beta_2}
		+
		\sqrt{\beta_1\beta_2\over (v/u)s}\,
		\right] 
		K_{1}\left(\sqrt{{uv}}\right)
		\Bigg\},
	\end{array}
	\label{main-pdf-1}
\end{align}
where $u=u_{0}$ and $v=v_{0}$, and $\boldsymbol{\theta}_0
=\boldsymbol{\theta}_{\rho=0}=(\alpha_1,\alpha_2,\beta_1,\beta_2,0)^\top$.
\begin{remark}
	Let $S=X/Y$ be the type II ratio, where $X\sim {\rm BS}(\alpha_1,\beta_1)$ and $Y\sim {\rm BS}(\alpha_2,\beta_2)$ are independent random variables (that is, $\rho=0$), and let $f_S(s;\boldsymbol{\theta}_0)$ denote its respective PDF.
	By using 
	the identity $f_S(s;\boldsymbol{\theta}_0)=\int_0^\infty y f_{X}(sy)f_{Y}(y) {\rm d}y$ it is simple to check that the PDF of $S$ is (for $s>0$)
	\begin{align}\label{PDF-S}
		\begin{array}{lll}
			f_S(s;\boldsymbol{\theta}_0)
			&=
			{\exp
				\left(
				{1\over\alpha_1^2}+{1\over\alpha_2^2}
				\right)\over 4\pi\alpha_1\alpha_2}\, 
			{1\over s}
			\\[0,5cm]
			&\times
			\Bigg\{
			\left(
			\sqrt{\beta_2 s\over\beta_1}
			+
			\sqrt{\beta_1\over \beta_2 s}
			\right) 
			K_{0}\left(\sqrt{{uv}}\right)
			+
			\left[
			\sqrt{(v/u)s\over\beta_1\beta_2}
			+
			\sqrt{\beta_1\beta_2\over (v/u)s}\,
			\right] 
			K_{1}\left(\sqrt{{uv}}\right)
			\Bigg\}
			\\[0,5cm]
			&=
			{1\over (s+1)^2}\, f_Z\left({1\over s+1};\boldsymbol{\theta}_0\right),
		\end{array}
	\end{align}
	where $f_Z(z;\boldsymbol{\theta}_0)$ is as given in \eqref{main-pdf-1}. The PDF \eqref{PDF-S} of $S$ with $\beta_1=\beta_2=1$ has appeared in reference  \cite{Teimouri2013}.
\end{remark}

\begin{figure}[htb!]
	\centering
	\includegraphics[width=1.0\linewidth]{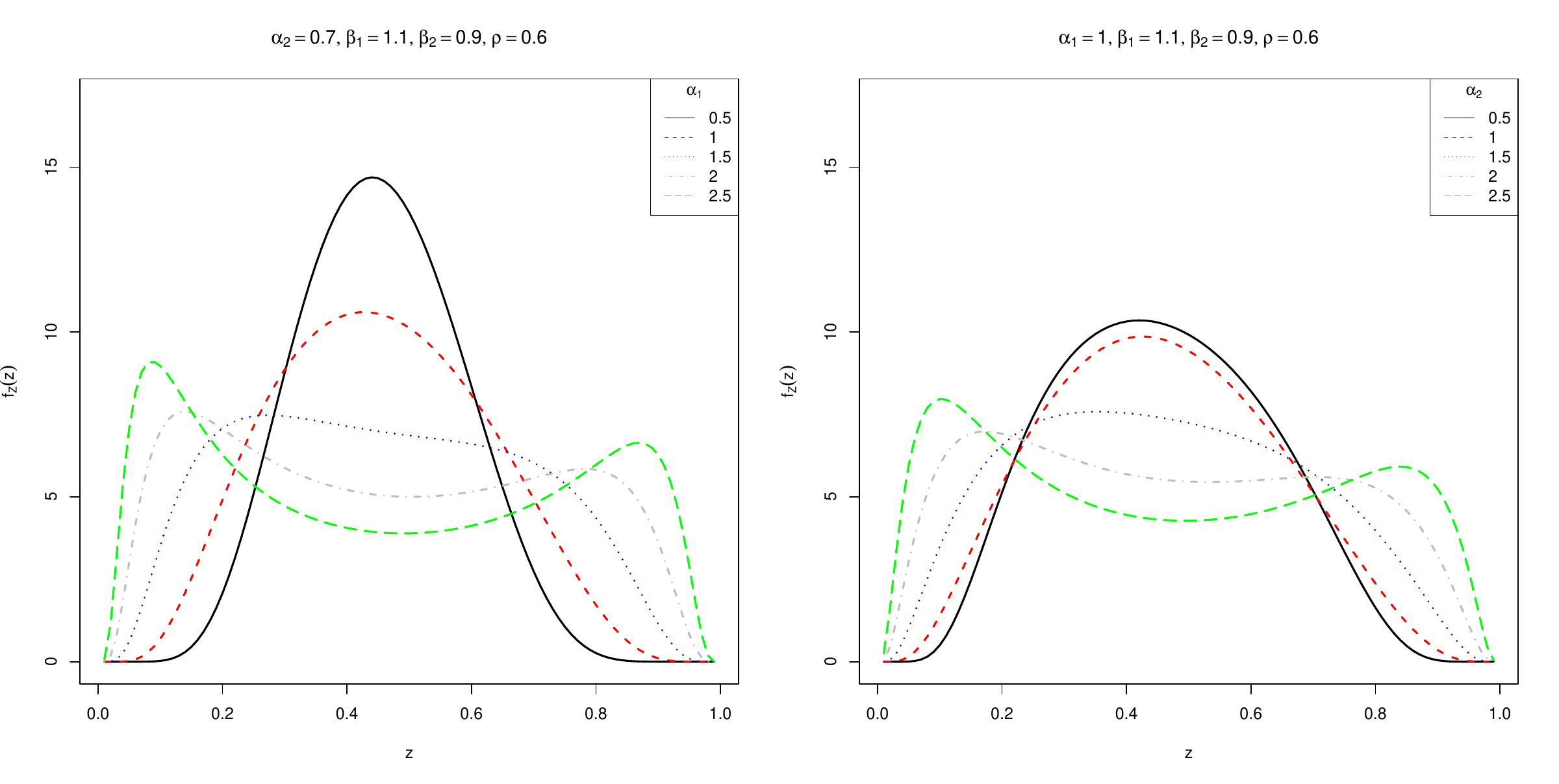}
	\caption{Plot of the PDF $f_Z$ with varying parameters $\alpha_1$ (left) and $\alpha_2$ (right).}
	\label{fig:f_Za}
\end{figure}

\begin{figure}[htb!]
	\centering
	\includegraphics[width=1.0\linewidth]{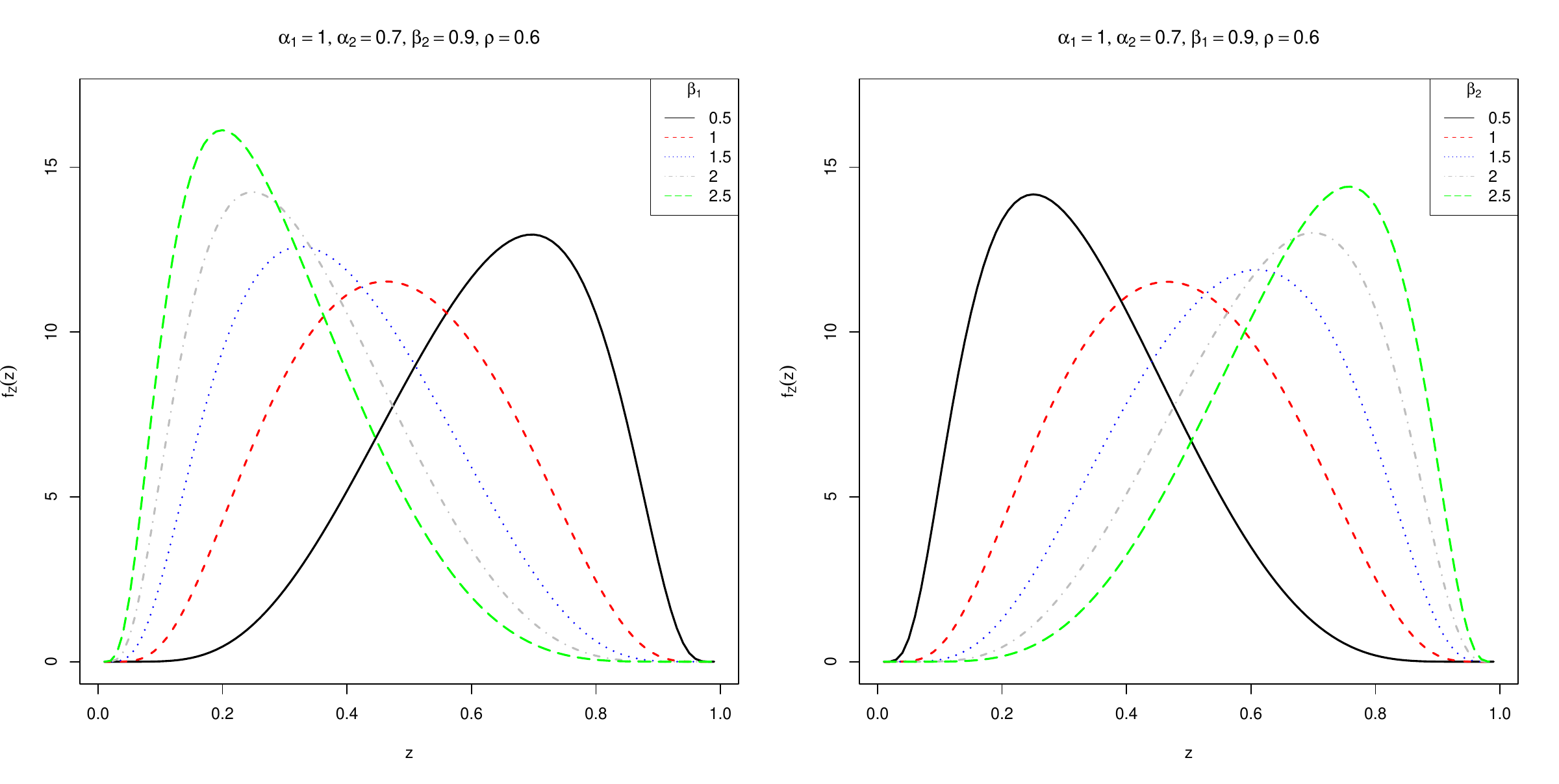}
	\caption{Plot of the PDF $f_Z$ with varying parameters $\beta_1$ (left) and $\beta_2$ (right).}
	\label{fig:f_Zb}
\end{figure}

\begin{figure}[htb!]
	\centering
	\includegraphics[width=1.0\linewidth]{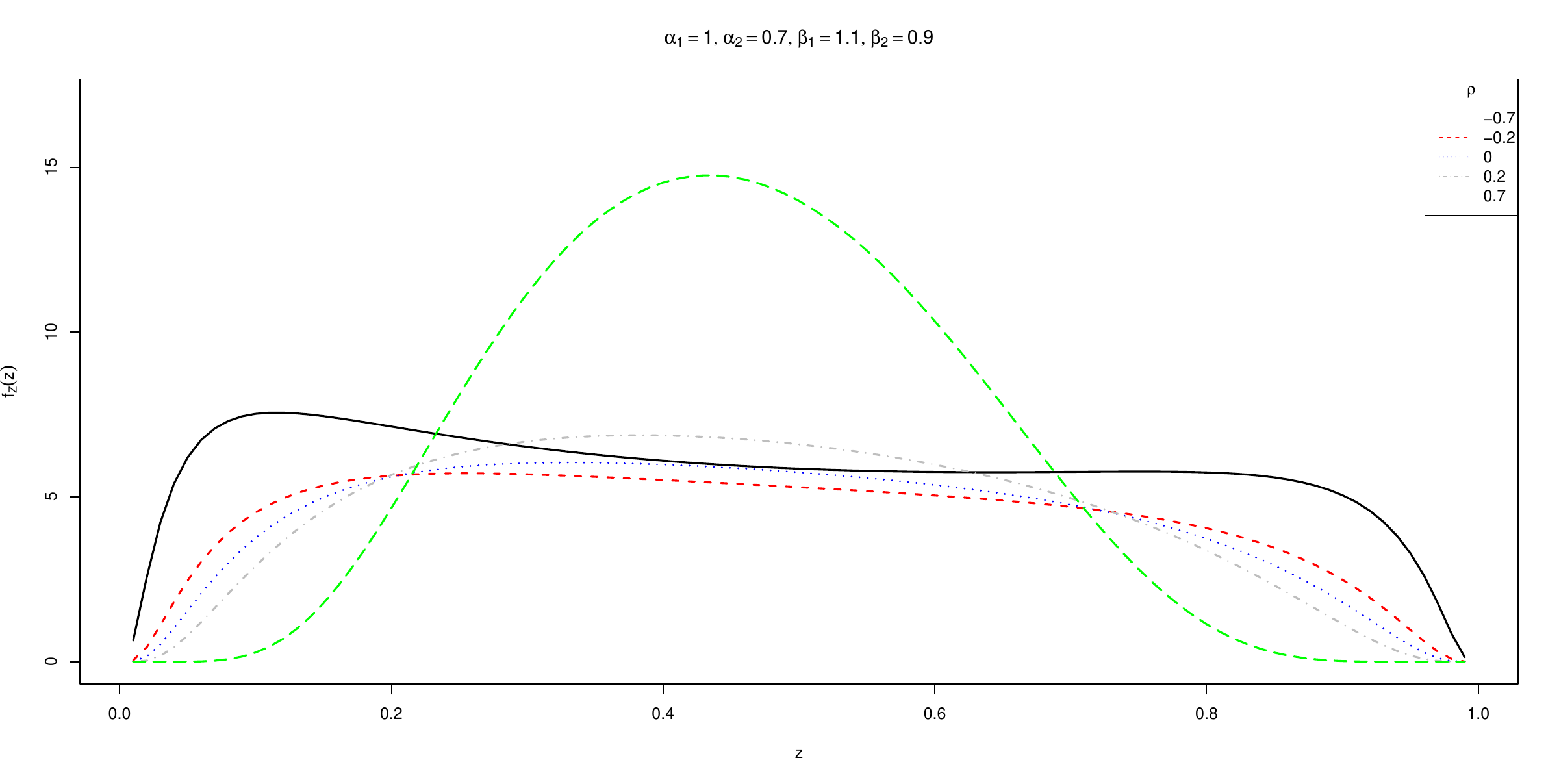}
	\caption{Plot of the PDF $f_Z$ with varying parameter $\rho$.}
	\label{fig:f_Zr}
\end{figure}

Figures \ref{fig:f_Za}, \ref{fig:f_Zb}, \ref{fig:f_Zr} and \ref{fig:pdf_same_beta} show the behavior of $f_Z$ for some parameter choices.
Note that the density modality changes according to the values of $\alpha_1$ and $\alpha_2$ parameters.
\textcolor{black}{Figure \ref{fig:pdf_same_beta} shows the symmetry of the distribution in the case where the scale parameters are equal. Figure \ref{fig:f_Zb} indicates that, for $\bm\theta_\rho = (\alpha_1, 0.7, 1.1, 0.9, 0.6)^\top$, the density exhibits a bimodal behavior when $\alpha_1 \geqslant 1.5$. A similar pattern is observed when $\alpha_1$ is fixed and $\alpha_2 \geqslant 2$.}


{\color{black}
	\begin{figure}[htb!]
		\centering
		\includegraphics[width=1.0\linewidth]{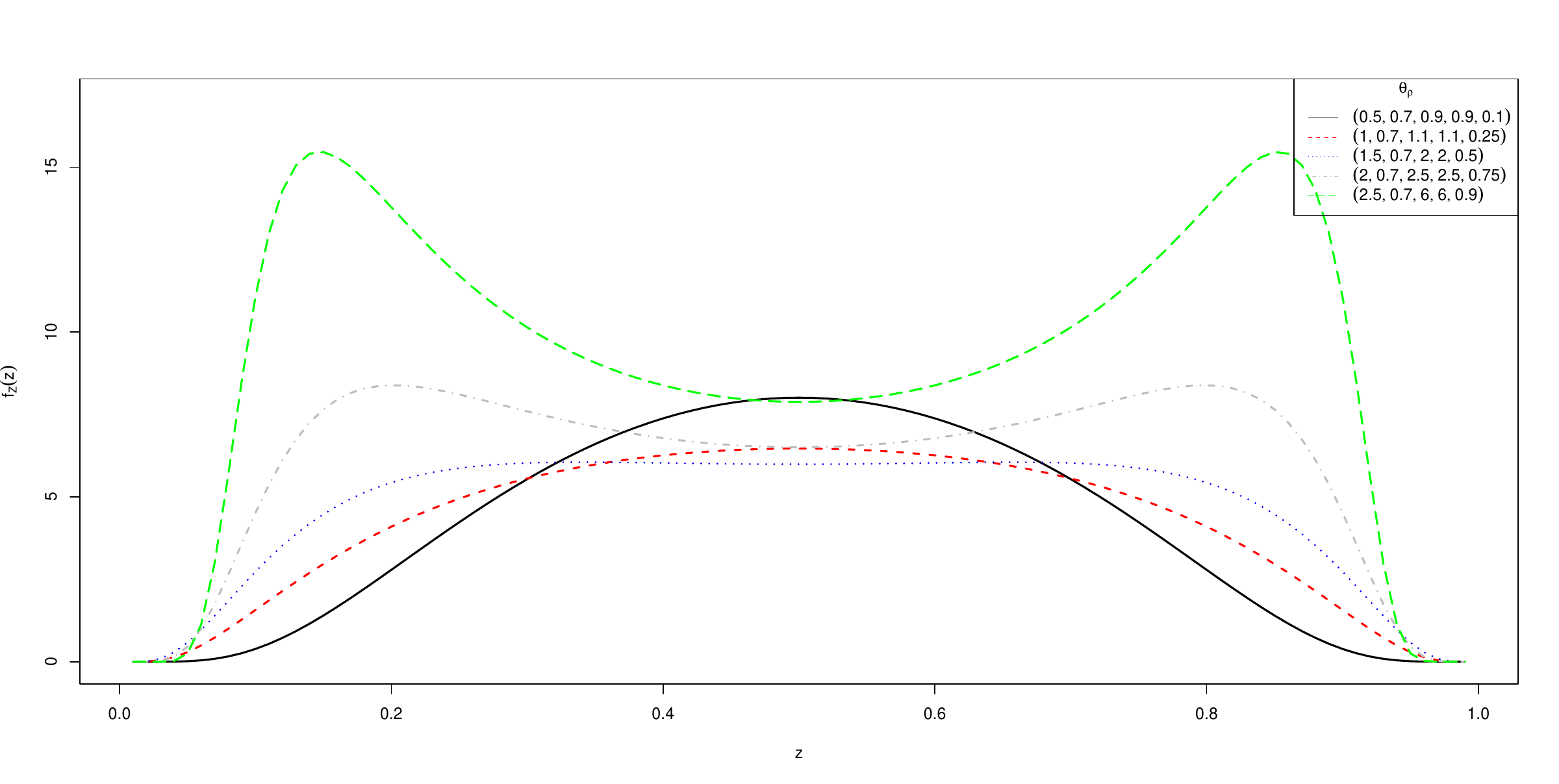}
		\caption{Plot of the PDF $f_Z$ with varying parameters $\beta_1=\beta_2$.}
		\label{fig:pdf_same_beta}
	\end{figure}	
}

{\color{black}
	\begin{remark}
		Note that the equation $f_Z'(z;\boldsymbol{\theta_\rho})=0$, where $f_Z'(z;\boldsymbol{\theta_\rho})$ is the derivative of $f_Z(z;\boldsymbol{\theta_\rho})$ in \eqref{main-pdf} with respect to $z$,  can equivalently be written as $G(z)=H(z)$, $0<z<1$, where $G$ and $H$ are non-polynomial functions (in terms of modified Bessel functions). In other words, the critical points of $f_Z$ are determined by the intersection points of (the graphs of the functions) $G$ and $H$. 
		The analysis of the intersection points of $G$ and $H$ can not be performed analytically, but graphical and numerical analyses are required. From Figures \ref{fig:f_Za}, \ref{fig:f_Zb} and \ref{fig:f_Zr} we come to the conclusion that the number of intersection points of $G$ and $H$ is bounded above by 3. That is, the function $f_Z$ has at most 3 critical points. Using this fact plus the fact that $\lim_{z\to 0^+}f_Z(z;\boldsymbol{\theta_\rho})=\lim_{z\to 1^-}f_Z(z;\boldsymbol{\theta_\rho})=0$ we conclude that $f_Z$ is unimodal (in the case that it has a single critical point) or bimodal (in the case of exactly 3 critical points). %Note that the limits $\lim_{z\to 0^+}f_Z(z;\boldsymbol{\theta_\rho})=\lim_{z\to 1^-}f_Z(z;\boldsymbol{\theta_\rho})=0$ plus the fact that $f_Z$ is a density rule out the no existence of critical points and the existence of 2 critical points for $f_Z$.
	\end{remark} 
}

\section{Characterization and estimation}\label{sec:2}

In this section, we establish some mathematical results of the {\color{black} UBBS1} distribution. In particular, we define the {\color{black} UBBS1} model as a ratio, present the cumulative distribution function{\color{black}, the moment-generating function and moments}, and calculate the stress-strength probability. We also derive parameter estimators for the {\color{black} UBBS1 model based on maximum likelihood and maximum product of spacings methods.}

% estimation

\subsection{{\color{black} UBBS1} model arising as a ratio}\label{UBS model arising as a ratio}
In this section, we establish one of the most important properties of the {\color{black} UBBS1} model. This property characterizes a {\color{black} UBBS1} random variable $Z$ in \eqref{main-pdf} as a ratio of the form $Y/(X+Y)$, where the random vector $(X,Y)^\top$ follows a bivariate Birnbaum-Saunders distribution as discussed in \cite{Kundu2017}.

Indeed, let $X, Y$ and $Z$ be continuous and positive random variables
such that
\begin{align}\label{main-id}
	Z\stackrel{d}{=}\dfrac{Y}{X+Y},
\end{align}
being $\stackrel{d}{=}$ equality in distribution, and the ratio in \eqref{main-id} has support in the unit interval $(0, 1)$. The random variable $Z = Y/(X+Y)$ is known in the literature as type I ratio \citep{Johnson95,Bekker2009}.

From \eqref{main-id}, it is simple to verify that the cumulative distribution function (CDF) of $Z$ is given by
\begin{align}\label{cdf}
	F_Z(z)
	=
	%\mathbb{P}\left({X\over X+Y}\leqslant z\right)
	%&=
	%=
	%\mathbb{P}\left({X\over Y}\leqslant {z\over 1-z}\right)
	%\\[0,2cm]
	%&=
	%=
	\mathbb{P}\left({X\over Y}\geqslant {s}\right)
	%=
	%\\[0,2cm]
	%&=
	%\mathbb{P}\left({Y\over X}\geqslant {s}\right)
	%\\[0,2cm]
	%&=
	%\mathbb{E}\left[\mathbb{P}\left({Y\over X}\geqslant {s}\Bigg\vert X\right)\right]
	%\\[0,2cm]
	%&=
	=
	1-
	\int_{0}^{\infty} \mathbb{P}\left(X\leqslant sy\vert Y=y\right) f_{Y}(y) {\rm d}y,
	\quad s={1\over z}-1.
\end{align}
%
%\begin{align*}
%f_Z(z)
%&=
%(s+1)^2
%\int_{0}^{\infty} x f_{Y}(sx \vert X=x) f_{X}(x) {\rm d}x
%\\[0,2cm]
%&=
%(s+1)^2
%\int_{0}^{\infty} x f_{Y,X}(sx,x) {\rm d}x
%\end{align*}

Hence, the	 PDF of $Z$ can be expressed as
\begin{align}\label{pdf-Z-0}
	f_Z(z)
	=
	(s+1)^2
	\int_0^\infty y f_{X,Y}(sy,y) {\rm d}y.
\end{align}

\begin{remark}
	Equation in \eqref{pdf-Z-0} can also be obtained using the Jacobian method by determining the marginal with respect to $Z$ of the random vector $(Z,U)^\top$, where $Z=Y/(X+Y)$ and $U=Y$.
\end{remark}

Now, let's suppose that $(X,Y)^\top$ has a bivariate Birnbaum-Saunders distribution with parameter vector $\boldsymbol{\theta}_{\rho}=(\alpha_1,\alpha_2,\beta_1,\beta_2,\rho)^\top$, denoted by $(X,Y)^\top\sim {\rm BS}(\boldsymbol{\theta})$, that is, its joint density is given by
\begin{align*}
	f_{X,Y}(x,y)
	=
	\phi_2(a(x;\alpha_1,\beta_1),a(y;\alpha_2,\beta_2);\rho)
	a'(x;\alpha_1,\beta_1)a'(y;\alpha_2,\beta_2),
\end{align*}
where $\phi_2(u, v;\rho)$ denotes the standard bivariate normal density function;
\begin{align*}
	\phi_2(u, v;\rho)
	=
	{1\over 2\pi\sqrt{1-\rho^2}}\,
	\exp\left\{-{1\over 2(1-\rho^2)}\, (u^2+v^2-2\rho uv)\right\},
	\quad (u,v)^\top\in\mathbb{R}^2
\end{align*}
and
\begin{align}\label{a-and-derivative}
	a(t;\alpha,\beta)
	=
	{1\over\alpha}
	\left(\sqrt{t\over\beta}-\sqrt{\beta\over t}\right),
	\quad 
	a'(t;\alpha,\beta)
	=
	{1\over 2\alpha t}
	\left(\sqrt{t\over\beta}+\sqrt{\beta\over t}\right),
	\quad t>0.
\end{align}

Hence, the density \eqref{pdf-Z-0} of $Z$ is written as
\begin{align}\label{pdf-Z}
	f_Z(z)
	=
	(s+1)^2
	\int_0^\infty y 
	\phi_2(a(sy;\alpha_1,\beta_1),a(y;\alpha_2,\beta_2);\rho)
	a'(sy,\alpha_1,\beta_1)a'(y,\alpha_2,\beta_2)
	{\rm d}y.
\end{align}

Since
\begin{align*}
	&a^2(sy;\alpha_1,\beta_1)+a^2(y;\alpha_2,\beta_2)-2\rho a(sy;\alpha_1,\beta_1)a(y;\alpha_2,\beta_2)
	\nonumber 
	\\[0,2cm]
	&=
	%{1\over\alpha_1^2}
	%\left({sy\over\beta_1}+{\beta_1\over sy}-2\right)
	%+
	%{1\over\alpha_2^2}
	%\left({y\over\beta_2}+{\beta_2\over y}-2\right)
	%-
	%{2\rho\over\alpha_1\alpha_2}
	%\left(
	%y\sqrt{s\over\beta_1\beta_2}
	%+
	%{1\over y}
	%\sqrt{\beta_1\beta_2\over s}
	%-
	%\sqrt{s\beta_2\over\beta_1}
	%-
	%\sqrt{\beta_1\over s\beta_2}
	%\right)
	%\\[0,2cm]
	%&=
	%\left(
	%{s\over\alpha_1^2\beta_1}
	%+
	%{1\over\alpha_2^2\beta_2}
	%-
	%{2\rho\over\alpha_1\alpha_2}
	%\sqrt{s\over\beta_1\beta_2}
	%\right)y
	%+
	%\left(
	%{\beta_1\over \alpha_1^2 s}
	%+
	%{\beta_2\over \alpha_2^2}
	%-
	%{2\rho\over\alpha_1\alpha_2}
	%\sqrt{\beta_1\beta_2\over s}
	%\right){1\over y}
	%-
	%2
	%\left[
	%{1\over\alpha_1^2}+{1\over\alpha_2^2}
	%-{\rho\over\alpha_1\alpha_2}\left(\sqrt{s\beta_2\over\beta_1}
	%+
	%\sqrt{\beta_1\over s\beta_2}\right)
	%\right]
	%\\[0,2cm]
	%&=
	%{y\over a_s}
	%+
	%{b_s\over y}
	%-
	%2
	%\left[
	%{1\over\alpha_1^2}
	%+
	%{1\over\alpha_2^2}
	%-{\rho\over\alpha_1\alpha_2}\left(\sqrt{s\beta_2\over\beta_1}
	%+
	%\sqrt{\beta_1\over s\beta_2}\right)
	%\right]
	%\\[0,2cm]
	%&=
	\left(
	{u_\rho y}
	+
	{v_\rho\over y}
	\right)
	-
	2
	\left[
	{1\over\alpha_1^2}
	+
	{1\over\alpha_2^2}
	-{\rho\over\alpha_1\alpha_2}\left(\sqrt{\beta_2 s\over\beta_1}
	+
	\sqrt{\beta_1\over \beta_2 s}\right)
	\right],
\end{align*}
with $u_\rho$ and $v_\rho$ as given in \eqref{def-a-b}, note that the function $\phi_2(a(sy;\alpha_1,\beta_1),a(y;\alpha_2,\beta_2);\rho)$ in \eqref{pdf-Z} can be expressed as
\begin{align}\label{identity-1-1}
	\phi_2(a(sy;\alpha_1,\beta_1),& \, a(y;\alpha_2,\beta_2);\rho)
	=
	{1\over 2}\,
	\exp\left\{-{1\over 2(1-\rho^2)}
	\left(
	{u_\rho y}
	+
	{v_\rho\over y}
	\right)\right\}
	\nonumber
	\\[0,2cm]
	&\times
	{1\over \pi\sqrt{1-\rho^2}}\, 
	\exp\left\{{1\over 1-\rho^2}
	\left[
	{1\over\alpha_1^2}
	+
	{1\over\alpha_2^2}
	-{\rho\over\alpha_1\alpha_2}\left(\sqrt{\beta_2 s\over\beta_1}
	+
	\sqrt{\beta_1\over\beta_2 s}\right)
	\right]\right\}.
\end{align}

Furthermore, notice that
\begin{align}\label{identity-2}
	a'(sy;\alpha_1,\beta_1)a'(y;\alpha_2,\beta_2)
	&=
	%{1\over 4\alpha_1\alpha_2 sy^2}
	%\left(\sqrt{sy\over\beta_1}+\sqrt{\beta_1\over sy}\right)
	%\left(\sqrt{y\over\beta_2}+\sqrt{\beta_2\over y}\right)
	%\\[0,2cm]
	%&=
	{1\over 4\alpha_1\alpha_2 sy}
	\left[
	{1\over y^2}
	\sqrt{\beta_1\beta_2\over s}
	+
	{1\over y}
	\left(
	\sqrt{\beta_2 s\over\beta_1}
	+
	\sqrt{\beta_1\over \beta_2 s}
	\right)
	+
	\sqrt{s\over\beta_1\beta_2}
	\right].
\end{align}

By replacing \eqref{identity-1-1} and \eqref{identity-2} in formula \eqref{pdf-Z}, we have
\begin{align}\label{pdf-Z-1}
	f_Z(z)
	&=
	{1\over \pi\sqrt{1-\rho^2}}\, 
	{(s+1)^2\over 4\alpha_1\alpha_2 s}\,
	\exp\left\{{1\over 1-\rho^2}
	\left[
	{1\over\alpha_1^2}
	+
	{1\over\alpha_2^2}
	-{\rho\over\alpha_1\alpha_2}\left(\sqrt{\beta_2 s\over\beta_1}
	+
	\sqrt{\beta_1\over \beta_2 s}\right)
	\right]\right\}
	\nonumber
	\\[0,2cm]
	&\times 
	\left[
	\sqrt{\beta_1\beta_2\over s}\,
	\mathbb{I}_1
	+
	\left(
	\sqrt{\beta_2 s\over\beta_1}
	+
	\sqrt{\beta_1\over \beta_2 s}
	\right) \mathbb{I}_2
	+
	\sqrt{s\over\beta_1\beta_2}\, \mathbb{I}_3
	\right],
\end{align}
where
\begin{align*}
	&\mathbb{I}_1
	=
	{1\over 2}\,
	\int_0^\infty 
	{1\over y^2}\,
	\exp\left\{-{1\over 2(1-\rho^2)}
	\left(
	{u_\rho y}
	+
	{v_\rho\over y}
	\right)\right\}
	{\rm d}y
	=
	\sqrt{u_\rho\over v_\rho}\, 
	K_{-1}\left({\sqrt{u_\rho v_\rho}\over 1-\rho^2}\right),
	\\[0,2cm]
	&\mathbb{I}_2
	=
	{1\over 2}\,
	\int_0^\infty 
	{1\over y}\,
	\exp\left\{-{1\over 2(1-\rho^2)}
	\left(
	{u_\rho y}
	+
	{v_\rho\over y}
	\right)\right\}
	{\rm d}y
	=
	K_{0}\left({\sqrt{u_\rho v_\rho }\over 1-\rho^2}\right)
\end{align*}
and
\begin{align*}
	\mathbb{I}_3
	=
	{1\over 2}\,
	\int_0^\infty 
	\exp\left\{-{1\over 2(1-\rho^2)}
	\left(
	{u_\rho y}
	+
	{v_\rho\over y}
	\right)\right\}
	{\rm d}y
	=
	\sqrt{v_\rho\over u_\rho}\, 
	K_{1}\left({\sqrt{{u_\rho v_\rho }}\over 1-\rho^2}\right).
\end{align*}
In the above, $K_\lambda(x)$, $x>0$, denotes the modified Bessel function of the third kind defined in \eqref{Bessel}.

Finally, by using that the Bessel function is symmetric with respect to the index $\lambda$: $K_\lambda(x)=K_{-\lambda}(x)$, and by employing the above formulas of $\mathbb{I}_1, \mathbb{I}_2$ and $\mathbb{I}_3$, from \eqref{pdf-Z-1} the PDF of $Z$ can be expressed as in \eqref{main-pdf}.

\subsection{The cumulative distribution function}\label{The cumulative distribution function}

For $(X,Y)^\top\sim {\rm BS}(\boldsymbol{\theta}_\rho)$, it holds that $X\sim {\rm BS}(\alpha_1,\beta_1)$ and $Y\sim {\rm BS}(\alpha_2,\beta_2)$. Furthermore, the  conditional CDF of $X$, given $Y=y$, is given by \citep[
Theorem 3.1]{Kundu2010}
\begin{align*}
	\mathbb{P}\left(X\leqslant x\vert Y=y\right) 
	=
	\Phi\left\{
	\dfrac{
		a(x;\alpha_1,\beta_1)
		-
		{\rho} a(y;\alpha_2,\beta_2)	
	}{\sqrt{1-\rho^2}}
	\right\},
\end{align*}
where $a(t;\alpha,\beta)$ is as given in \eqref{a-and-derivative} and $\Phi$ is the CDF of the standard normal distribution.
Hence, from \eqref{cdf} it follows that
\begin{align*}
	F_Z(z;\boldsymbol{\theta}_\rho)
	&=
	1-
	\int_{0}^{\infty} 
	\mathbb{P}\left(X\leqslant sy\vert Y=y\right) 
	f_{Y}(y) {\rm d}y,
	\quad s={1\over z}-1,
	\nonumber
	\\[0,2cm]
	&=
	1-
	\int_{0}^{\infty} 
	\Phi\left\{
	\dfrac{
		a(sy;\alpha_1,\beta_1)
		-
		{\rho}a(y;\alpha_2,\beta_2)
	}{\sqrt{1-\rho^2}}
	\right\}
	\phi(a(y;\alpha_2,\beta_2))a'(y;\alpha_2,\beta_2)
	{\rm d}y,
	%\\[0,2cm]
	%&=
	%1-\mathbb{E}\left[\Phi\left\{
	%\dfrac{
		%	a(sY;\alpha_1,\beta_1)
		%	-
		%	{\rho}a(Y;\alpha_2,\beta_2)
		%}{\sqrt{1-\rho^2}}
	%\right\}\right],
	%\quad Y\sim {\rm BS}(\alpha_2,\beta_2),
	%\nonumber
\end{align*}
where $\phi$ denotes the PDF of the standard normal distribution.
By using the change of variable $w=a(y;\alpha_2,\beta_2)$, we write the above expression as

\begin{align*}
	&1-
	\int_{-\infty}^{\infty} 
	\Phi\left\{
	\dfrac{
		a(sa^{-1}(w;\alpha_2,\beta_2);\alpha_1,\beta_1)
		-
		{\rho}w
	}{\sqrt{1-\rho^2}}
	\right\}
	\phi(w)
	{\rm d}w
	\\[0,2cm]
	&=
	{1\over 2}
	-
	{1\over 2}
	\int_{-\infty}^{\infty} 
	{\rm erf}\left\{
	\dfrac{
		a(sa^{-1}(w;\alpha_2,\beta_2);\alpha_1,\beta_1)
		-
		{\rho}w
	}{\sqrt{2(1-\rho^2)}}
	\right\}
	\phi(w)
	{\rm d}w,
\end{align*}
where we have used the well-known identity 
$\Phi(x)=({1}/{2})[1+{\mbox{erf}}(x/\sqrt{2})]$. In the above, $a^{-1}(t;\alpha,\beta)$ denotes the inverse function of $a(t;\alpha,\beta)$.
As 
$
a^{-1}(t;\alpha,\beta)
=
(\beta/4)[\alpha t+\sqrt{(\alpha t)^2+4}\, ]^2,
$
the integral above can be written as
\begin{align*}
	{1\over 2}
	-\!
	{1\over 2}
	\int_{-\infty}^{\infty} 
	{\rm erf}\!\left\{\!
	{
		{\left({\alpha_2\over\alpha_1}\sqrt{s\beta_2\over\beta_1}-\rho\right)\over \sqrt{2(1-\rho^2)}}\,
		w
		\!+\!
		{\sqrt{2}(s\beta_2-\beta_1)\over \alpha_1\sqrt{s\beta_1\beta_2(1-\rho^2)}}
		{1\over \alpha_2w+\sqrt{(\alpha_2 w)^2+4}}
	}\!
	\right\}\!
	\phi(w)
	{\rm d}w.
\end{align*}

Therefore, the CDF of the {\color{black} UBBS1} random variable $Z$ is given by (for $s={1/z}-1$ and $0<z<1$)
\begin{align}\label{cdf-10}
	\begin{array}{lll}
		F_Z(z;\boldsymbol{\theta}_\rho)
		=\!
		{1\over 2}
		-
		{1\over 2}
		{\displaystyle
			\int_{-\infty}^{\infty} 
		}\!
		{\rm erf}\!\left\{
		{
			{\left({\alpha_2\over\alpha_1}\sqrt{s\beta_2\over\beta_1}-\rho\right)\over \sqrt{2(1-\rho^2)}}\,
			w
			+
			{\sqrt{2}(s\beta_2-\beta_1)\over \alpha_1\sqrt{s\beta_1\beta_2(1-\rho^2)}}\,
			{1\over \alpha_2w+\sqrt{(\alpha_2 w)^2+4}}
		}
		\right\}\!
		\phi(w)
		{\rm d}w.
	\end{array}
\end{align}

{\color{black}
	\subsection{Moment-generating function and moments}\label{Moment-generating function and moments}
	
	Let $Z\sim {\rm UBBS1}(\boldsymbol{\theta}_\rho)$. Since $Z$ has unitary support it is clear that the moment-generating function (MGF), denoted by $M_Z(t)=\mathbb{E}[\exp(tZ)]$, $t\in\mathbb{R}$, and the (raw) moments, denoted by $\mu_n=\mathbb{E}(Z^n)$, $n\in\mathbb{N}$, are finite. In particular, $Z$ has a finite MGF $M_Z(t)$
	in some open interval containing $0$, then it holds $M_Z^{(n)}(t)\big\vert_{t=0}=\mu_n$, and $M_Z(t)$
	may be expanded in a power series about $0$ as follows
	\begin{align*}
		M_Z(t)=1+\sum_{n=1}^{\infty} {\mu_n\over n!}\, t^n.
	\end{align*}
	
	On the other hand, by using the well-known formula $\mathbb{E}(X^p)=p\int_0^\infty x^{p-1}\mathbb{P}(X>x){\rm d}x$, $X>0$, $p>0$, we get
	\begin{align*}
		\mu_n
		=
		n\int_0^1 z^{n-1}[1-	F_Z(z;\boldsymbol{\theta}_\rho)]{\rm d}z,
	\end{align*}
	where $	F_Z(z;\boldsymbol{\theta}_\rho)$ is as given in \eqref{cdf-10}. Hence, from \eqref{cdf-10},
	\begin{multline}\label{moments}
		\mu_n
		=
		%	n\int_0^1 z^{n-1}[1-	F_Z(z;\boldsymbol{\theta}_\rho)]{\rm d}z
		%	\\[0,1cm]
		%	&=
		{1\over 2}
		+
		{n\over 2}
		\int_0^1 
		z^{n-1}
		\left[
		{\displaystyle
			\int_{-\infty}^{\infty} 
		}
		{\rm erf}\left\{
		{\big(\scriptstyle {\alpha_2\over\alpha_1}\sqrt{s\beta_2\over\beta_1}-\rho\big)\over \scriptstyle\sqrt{2(1-\rho^2)}}\,
		w
		\right.
		\right.
		\\[0,1cm]	
		+
		\left.
		\left.
		{\sqrt{2}(s\beta_2-\beta_1)\over \alpha_1\sqrt{s\beta_1\beta_2(1-\rho^2)}}\,
		{1\over \alpha_2w+\sqrt{(\alpha_2 w)^2+4}}
		\right\}
		\phi(w)
		{\rm d}w\right]
		{\rm d}z.
	\end{multline}
	In the special case  that $\beta_1=\beta_2$, we have $\mu_1=1/2$ (see Remark \ref{rem-1}). In general, the integral in \eqref{moments} cannot be solved analytically, 
	%nor expressed in terms of known mathematical functions. 
	so the use of numerical method is essential for its respective calculation.
	
	Table \ref{correlationstable} shows some values of moments (with $n=1,\ldots,10,50,100$) in \eqref{moments} for different choices of $\rho$ in the special case when $\alpha_1=\alpha_2=\beta_1=\beta_2=1$.

			\begin{table}[ht]
				\centering
				\begin{adjustbox}{width=1\textwidth}
					\small
					\begin{tabular}{lrrrrrrrrrrrrrrrrrrr}
						\hline
						$\rho$ &-0.900 & -0.750 & -0.500 & -0.250 & -0.100 & 0.000 & 0.100 & 0.250 & 0.500 & 0.750 & 0.900
						\\ 		\hline
						$\mu_1$&0.500& 0.500& 0.500& 0.500& 0.500& 0.500& 0.500& 0.500& 0.500& 0.500& 0.500\\
						$\mu_2$&0.259& 0.269& 0.279& 0.284& 0.285& 0.286& 0.285& 0.284& 0.279& 0.269& 0.259\\
						$\mu_3$&0.139& 0.153& 0.168& 0.176& 0.178& 0.178& 0.178& 0.176& 0.168& 0.153& 0.139\\
						$\mu_4$&0.076& 0.092& 0.108& 0.116& 0.118& 0.119& 0.118& 0.116& 0.108& 0.092& 0.076\\
						$\mu_5$&0.043& 0.057& 0.072& 0.080& 0.082& 0.083& 0.082& 0.080& 0.072& 0.057& 0.043\\
						$\mu_6$&0.025& 0.037& 0.050& 0.057& 0.059& 0.060& 0.059& 0.057& 0.050& 0.037& 0.025\\
						$\mu_7$& 0.015& 0.024& 0.036& 0.042& 0.044& 0.044& 0.044& 0.042& 0.036& 0.024& 0.015\\
						$\mu_8$&0.009& 0.016& 0.026& 0.032& 0.033& 0.034& 0.033& 0.032& 0.026& 0.016& 0.009\\
						$\mu_9$&0.005& 0.011& 0.019& 0.024& 0.026& 0.026& 0.026& 0.024& 0.019& 0.011& 0.005\\
						$\mu_{10}$&0.003& 0.008& 0.015& 0.019& 0.020& 0.020& 0.020& 0.019& 0.015& 0.008& 0.003\\
						$\mu_{50}$&$<10^{-7}$& $<10^{-5}$& $<10^{-4}$& $<10^{-4}$& $<10^{-4}$& $<10^{-4}$& $<10^{-4}$& $<10^{-4}$& $<10^{-4}$ &$<10^{-5}$& $<10^{-7}$\\
						$\mu_{100}$&$<10^{-11}$& $<10^{-7}$& $<10^{-6}$& $<10^{-5}$&$<10^{-5}$& $<10^{-5}$& $<10^{-5}$& $<10^{-5}$& $<10^{-6}$& $<10^{-7}$& $<10^{-11}$\\
						\hline
					\end{tabular}
				\end{adjustbox}
				\caption{\color{black}Moments values for different choices of $\rho$.}\label{correlationstable}
			\end{table}

		}
		
		\subsection{The stress-strength probability}\label{The stress-strength probability}
		
		%By using the assumption $\beta_1=\beta_2$, 
		In what follows we explicitly calculate the stress-strength probability $R\equiv
		\mathbb{P}(X<Y)$  in the case where $(X,Y)^\top\sim {\rm BS}(\boldsymbol{\theta}_\rho)$, that is,
		$X$ and $Y$ can admit correlation. 
		
		Indeed, by \eqref{cdf}, we have
		\begin{align*}
			R
			=
			\mathbb{P}\left({X\over Y}< 1\right)
			=
			1-\mathbb{P}\left({X\over Y}\geqslant 1\right)
			=
			1-F_Z(z;\boldsymbol{\theta}_\rho)\Big\vert_{z=1/2}.
		\end{align*}
		Hence, as $z=1/2$ if and only if $s=1$, by using \eqref{cdf-10},
		% with $\beta_1=\beta_2$ and $s=1$, 
		we get
		
		\begin{align}\label{stress-strength}
			\begin{array}{lll}
				R
				=
				{1\over 2}
				+
				{1\over 2}
				{
					\displaystyle
					\int_{-\infty}^{\infty} 
				}
				{\rm erf}\left\{
				{
					{\left({\alpha_2\over\alpha_1}\sqrt{\beta_2\over\beta_1}-\rho\right)\over \sqrt{2(1-\rho^2)}}\,
					w
					+
					{\sqrt{2}(\beta_2-\beta_1)\over \alpha_1\sqrt{\beta_1\beta_2(1-\rho^2)}}\,
					{1\over \alpha_2w+\sqrt{(\alpha_2 w)^2+4}}
				}
				\right\}
				\phi(w)
				{\rm d}w.
			\end{array}
		\end{align}
		
		In the particular case $\beta_1=\beta_2$, from \eqref{stress-strength}, the stress-strength probability $R$ is simply written as
		\begin{align*}
			R
			=
			{1\over 2}
			+
			{1\over 2}
			\int_{-\infty}^{\infty} 
			{\rm erf}\left\{
			\dfrac{
				\left(
				{\alpha_2\over\alpha_1}
				-
				{\rho}
				\right)
			}{\sqrt{2(1-\rho^2)}}\,
			w
			\right\}
			\phi(w) {\rm d}w
			=
			{1\over 2},
			\quad 
			\forall \rho\in(-1,1),
		\end{align*}
		where in the last equality we have used the fact that erf$(x)$ and $\phi$ are odd and even functions, respectively.
		
		To our knowledge, in this setting ($\beta_1=\beta_2$), identity $R=1/2$ was known only for the special case in which $X$ and $Y$ are independent (that is, $\rho=0$); see, e.g. \cite{Saulo2019} and \cite{Quintino2024}.

		{\color{black}
			
			\subsubsection{Maximum likelihood estimation}\label{Maximum likelihood estimation}
			
			Let $\{Z_i:i = 1,\ldots,n\}$ be a univariate random sample of size $n$ from the UBS$(\boldsymbol{\theta}_\rho)$
			distribution with PDF as in \eqref{main-pdf}, and let $z_i$ be the sample observation of $Z_i$. Then,
			the log-likelihood function for $\boldsymbol{\theta}_\rho=(\alpha_1,\alpha_2,\beta_1,\beta_2,\rho)^\top$, $\alpha_1> 0,\alpha_2> 0 ,\beta_1> 0, \beta_2> 0$ and $\rho\in(-1,1)$, is given by (without additive constant)
			\begin{align*}
				\begin{array}{lll}
					\ell(\boldsymbol{\theta}_\rho)
					&=
					{n\over 1-\rho^2}
					\left(
					{1\over\alpha_1^2}+{1\over\alpha_2^2}
					\right)
					-
					n
					\log\left(4\alpha_1\alpha_2 \pi\sqrt{1-\rho^2}\right)
%					\\[0,4cm]
%					&
					-
					{\rho\over \alpha_1\alpha_2(1-\rho^2)}
					\sum_{i=1}^{n}
					\left(\sqrt{\beta_2 s_i\over\beta_1}
					+
					\sqrt{\beta_1\over \beta_2 s_i}\right)
					\nonumber
					\\[0,4cm]
					&
					+
					\sum_{i=1}^{n}
					\log
					\Bigg\{
					\left(
					\sqrt{\beta_2 s_i\over\beta_1}
					+
					\sqrt{\beta_1\over\beta_2 s_i}
					\right)
					K_{0}\left({\sqrt{u_{\rho,i} v_{\rho,i}}\over 1-\rho^2}\right)
%					\\[0,4cm]
%					&
%					\hspace*{4cm}
					+
					\left[
					\sqrt{(v_{\rho,i}/u_{\rho,i}) s_i\over\beta_1\beta_2}
					+
					\sqrt{\beta_1\beta_2\over  (v_{\rho,i}/u_{\rho,i}) s_i}\,
					\right]
					K_{1}\left({\sqrt{u_{\rho,i}v_{\rho,i}}\over 1-\rho^2}\right)
					\Bigg\},
				\end{array}
			\end{align*}
			where
			\begin{align*}
				&s_i={1\over z_i}-1,
				\quad i=1,\ldots,n,
				\\[0,2cm]
				&u_{\rho,i}
				\equiv
				{s_i\over\alpha_1^2\beta_1}
				+
				{1\over\alpha_2^2\beta_2}
				-
				{2\rho\over\alpha_1\alpha_2}
				\sqrt{s_i\over\beta_1\beta_2},
				\quad
				v_{\rho,i}
				\equiv
				{\beta_1\over \alpha_1^2 s_i}
				+
				{\beta_2\over \alpha_2^2}
				-
				{2\rho\over\alpha_1\alpha_2}
				\sqrt{\beta_1\beta_2\over s_i},
			\end{align*}
			and
			$K_\lambda(x)$, $x>0$, is the modified Bessel function of the third kind defined in \eqref{Bessel}.
			
			The necessary conditions for the
			occurrence of a maximum (or a minimum) are
			\begin{align*}
				{\partial \ell(\boldsymbol{\theta}_\rho)\over\partial\alpha_1}=0,
				\quad
				{\partial \ell(\boldsymbol{\theta}_\rho)\over\partial\alpha_2}=0,
				\quad
				{\partial \ell(\boldsymbol{\theta}_\rho)\over\partial\beta_1}=0,
				\quad
				{\partial \ell(\boldsymbol{\theta}_\rho)\over\partial\beta_2}=0,
				\quad
				{\partial \ell(\boldsymbol{\theta}_\rho)\over\partial\rho}=0,
			\end{align*}
			known as the likelihood equations, with
			%where we used the fact that $\partial K_\nu(x)/\partial x=-K_{\nu-1}(x)-(\nu/x)K_\nu(x)$.
			\begin{align*}
				\begin{array}{lll}
					&{\partial \ell(\boldsymbol{\theta}_\rho)\over\partial\alpha_1}
					=
					-
					\frac{2 n}{\left(1-\rho^{2}\right) \alpha_{1}^{3}}-\frac{n}{\alpha_{1}}+\frac{\rho}{\alpha_{1}^{2} \alpha_{2}\left(1-\rho^{2}\right)} \sum_{i=1}^{n}\left(\sqrt{\frac{ \beta_{2}s_{i}}{\beta_{1}}}+\sqrt{\frac{\beta_{1}}{ \beta_{2}s_{i}}}\right)+\sum_{i=1}^{n}
					{{\partial G(\bm\theta_\rho, s_i)\over\partial \alpha_{1}}\over G(\bm\theta_\rho, s_i)},
					\\[0,4cm]
					&{\partial \ell(\boldsymbol{\theta}_\rho)\over\partial\alpha_2}
					=
					-
					\frac{2 n}{\left(1-\rho^{2}\right) \alpha_{2}^{3}}-\frac{n}{\alpha_{2}}+\frac{\rho}{\alpha_{1} \alpha_{2}^{2}\left(1-\rho^{2}\right)} \sum_{i=1}^{n}\left(\sqrt{\frac{ \beta_{2}s_{i}}{\beta_{1}}}+\sqrt{\frac{\beta_{1}}{ \beta_{2} s_{i}}}\right)+\sum_{i=1}^{n}
					{{\partial G(\bm\theta_\rho, s_i)\over \partial \alpha_{2}}\over  G(\bm\theta_\rho, s_i)},
					\\[0,4cm]
					&{\partial \ell(\boldsymbol{\theta}_\rho)\over\partial\beta_1}
					=-\frac{\rho}{2\alpha_{1} \alpha_{2}\left(1-\rho^{2}\right)} \sum_{i=1}^{n}\left(\frac{1}{\sqrt{\beta_{1}\beta_{2} s_{i}}}- \sqrt{\frac{ \beta_{2}s_{i}}{\beta_{1}^{3}}}\right)+
					\sum_{i=1}^{n}
					{{\partial G(\bm\theta_\rho, s_i)\over \partial \beta_{1}}\over  G(\bm\theta_\rho, s_i)},
					\\[0,4cm]
					&{\partial \ell(\boldsymbol{\theta}_\rho)\over\partial\beta_2}
					=-\frac{\rho}{2\alpha_{1} \alpha_{2}\left(1-\rho^{2}\right)} \sum_{i=1}^{n}\left(\frac{1}{\sqrt{\beta_{1}\beta_{2} s_{i}}}- \sqrt{\frac{\beta_{1}}{ \beta_{2}^{3}s_{i}^{3}}}\right) s_{i}+\sum_{i=1}^{n} {{\partial\log G(\bm\theta_\rho, s_i)\over \partial \beta_{2}}\over  G(\bm\theta_\rho, s_i)},
					\\[0,4cm]
					&{\partial \ell(\boldsymbol{\theta}_\rho)\over\partial\rho}
					=\frac{2 n \rho}{\left(1-\rho^{2}\right)^{2}}\left(\frac{1}{\alpha_{1}^{2}}+\frac{1}{\alpha_{2}^{2}}\right)-
					\frac{n\rho}{1-\rho^{2}}-\frac{1+\rho^{2}}{\alpha_{1} \alpha_{2}\left(1-\rho^{2}\right)^{2}} \sum_{i=1}^{n}\left(\sqrt{\frac{\beta_{2} s_{i}}{\beta_{1}}}+\sqrt{\frac{\beta_{1}}{ \beta_{2} s_{i}}}\right)
%					\\[0,4cm]
%					&
%					\hspace{9.5cm}
					+\sum_{i=1}^{n}
					{{\partial G\left(\bm\theta_\rho, s_{i}\right)\over \partial \rho}\over  G(\bm\theta_\rho, s_i)},
				\end{array}
			\end{align*}
			where we denote
			\begin{equation*}
				\begin{array}{lll}
					G(\bm\theta_\rho, s_i)
					&=
					\left(\sqrt{\frac{\beta_{2 }s_i}{\beta_{1}}}+\sqrt{\frac{\beta_{1}}{\beta_{2} s_i}}\right)
					K_{0}\left(\frac{\sqrt{u_{\rho,i}v_{\rho,i}}}{1-\rho^{2}}\right)
%					\\[0,4cm]
%					&		 
+
					\left[
					\sqrt{\frac{(v_{\rho,i}/u_{\rho,i}) s_i}{\beta_{1} \beta_{2}}}+\sqrt{\frac{\beta_{1} \beta_{2}}{ (v_{\rho,i}/u_{\rho,i}) s_i}}\,
					\right]
					K_{1}\left(
					\frac{\sqrt{u_{\rho,i}v_{\rho,i}}}{1-\rho^{2}}
					\right),
				\end{array}
			\end{equation*}
			and its partial derivatives are obtained using the known fact that $\partial K_\nu(x)/\partial x=-K_{\nu-1}(x)-(\nu/x)K_\nu(x)$.
			
			Note that
			\begin{align*}
				\begin{array}{lll}
					G\left(\bm\theta_\rho, s_i\right)
					&=
					\beta_{2} s_i\,
					a'\left(\beta_{2} s_i; {1\over 2}, \beta_{1}\right) K_{0}\left(\frac{\sqrt{u_{\rho,i}v_{\rho,i}}}{1-\rho^{2}} \right)
%					\\[0,4cm]
%					&
					+
					{v_{\rho,i}\over u_{\rho,i}}\, s_i\,
					a'\left({v_{\rho,i}\over u_{\rho,i}}\,  s_i; {1\over 2}, \beta_{1} \beta_{2}\right)
					K_{1}\left(\frac{\sqrt{u_{\rho,i}v_{\rho,i}}}{1-\rho^{2}} \right),
				\end{array}
			\end{align*}
			where $a'(t, \alpha, \beta)$ is given in \eqref{a-and-derivative}.
			
			We require additional partial derivatives of $G(\bm{\theta}_\rho, s_i)$. Essentially, the product and chain rules are employed for deriving the subsequent functions:
			\begin{enumerate}[(a)]
				\item Partial derivatives with respect to $\alpha_{1}$:
				\begin{align*}
					\begin{array}{lll}
						\frac{\partial v_{\rho,i}}{\partial \alpha_{1}}
						=
						- \frac{2\beta_{1}}{\alpha_{1}^{3} s_i}+\frac{2 \rho}{\alpha_{1}^{2} \alpha_{2}} \sqrt{\frac{\beta_{1} \beta_{2}}{s_i}},
						\quad
						\frac{\partial u_{\rho,i}}{\partial \alpha_{1}}
						=
						-
						\frac{2s_i}{\alpha_{1}^{3}\beta_{1}}
						+
						\frac{2 \rho}{\alpha_{1}^{2} \alpha_{2}} \sqrt{\frac{s_i}{\beta_{1} \beta_{2}}};
					\end{array}
				\end{align*}
				
				\item Partial derivatives with respect to $\alpha_{2}$:
				\begin{align*}
					\begin{array}{lll}
						\frac{\partial v_{\rho,i}}{\partial \alpha_{2}}=-\frac{2 \beta_{2}}{\alpha_{2}^{3}}+\frac{2 \rho}{\alpha_{1} \alpha_{2}^{2}} \sqrt{\frac{\beta_{1} \beta_{2}}{s_i}},
						\quad
						\frac{\partial u_{\rho,i}}{\partial \alpha_{2}}=
						-\frac{2}{\alpha_{2}^{3} \beta_{2}}+\frac{2 \rho}{\alpha_{1} \alpha_{2}^{2}} \sqrt{\frac{\beta_{1} \beta_{2}}{s_i}};
					\end{array}
				\end{align*}
				
				\item Partial derivatives with respect to $\beta_1$:
				\begin{align*}
					\begin{array}{lll}
						\frac{\partial v_{\rho,i}}{\partial \beta_{1}}
						=
						\frac{1}{\alpha_{1}^{2} s_i}-\frac{\rho}{\alpha_{1} \alpha_{2}} \sqrt{\frac{\beta_{2}}{\beta_{1} s_i}}, \quad
						\frac{\partial u_{\rho,i}}{\partial \beta_{1}}
						=
						-{s_i\over\alpha_1^2\beta_1^2}
						+
						{\rho\over\alpha_1\alpha_2}
						\sqrt{s_i\over\beta_1^3\beta_2}
						;
					\end{array}
				\end{align*}
				
				\item Partial derivatives with respect to $\beta_{2}$:
				\begin{align*}
					\begin{array}{lll}
						\frac{\partial v_{\rho,i}}{\partial \beta_{2}}=\frac{1}{\alpha_{2}^{2}}-\frac{\rho}{\alpha_{1} \alpha_{2}} \sqrt{\frac{\beta_{1}}{\beta_{2} s_i}},
						\quad
						\frac{\partial u_{\rho,i}}{\partial \beta_{2}}
						=
						-
						\frac{1}{\alpha_{2}^{2} \beta_{2}^{2}}
						+
						\frac{\rho}{\alpha_{1} \alpha_{2}} \sqrt{\frac{s_i}{\beta_{1} \beta_{2}^{3}}};
					\end{array}
				\end{align*}
				
				\item Partial derivatives with respect to $\rho$:
				\begin{align*}
					\begin{array}{lll}
						\frac{\partial v_{\rho,i}}{\partial \rho}
						=
						-\frac{2}{\alpha_{1} \alpha_{2}} \sqrt{\frac{\beta_{1} \beta_{2}}{s_i}},
						\quad
						\frac{\partial u_{\rho,i}}{\partial \rho}
						=
						-
						\frac{2}{\alpha_{1} \alpha_{2}} \sqrt{\frac{s_i}{\beta_{1} \beta_{2}}}.
					\end{array}
				\end{align*}
			\end{enumerate}
			Finally, for $\theta_{k} \in\left\{\alpha_{1}, \alpha_{2}, \beta_{1}, \beta_{2}, \rho\right\},$ we have
			\begin{equation*}
				\begin{array}{lll}
					\frac{\partial}{\partial \theta_{k}} K_{0}\left(\frac{\sqrt{u_{\rho,i}v_{\rho,i}}}{1-\rho^{2}}\right)
					=
					-K_{-1}\left(\frac{\sqrt{u_{\rho,i}v_{\rho,i}}}{1-\rho^{2}}\right) \frac{\partial}{\partial \theta_{k}}\left(\frac{\sqrt{u_{\rho,i}v_{\rho,i}}}{1-\rho^{2}}\right),
				\end{array}
			\end{equation*}
			and
			\begin{align*}
				\begin{array}{lll}
					\nonumber \frac{\partial}{\partial \theta_{k}} K_{1}\left(\frac{\sqrt{u_{\rho,i}v_{\rho,i}}}{1-\rho^{2}} \right)
					=
					\left[-K_{0}\left(\frac{\sqrt{u_{\rho,i}v_{\rho,i}}}{1-\rho^{2}} \right)-\left(\frac{\sqrt{u_{\rho,i}v_{\rho,i}}}{1-\rho^{2}}\right)^{-1} K_{1}\left(\frac{\sqrt{u_{\rho,i}v_{\rho,i}}}{1-\rho^{2}} \right)\right]\!
					\frac{\partial}{\partial \theta_{k}}\left(\frac{\sqrt{u_{\rho,i}v_{\rho,i}}}{1-\rho^{2}}\right).
				\end{array}
			\end{align*}
			
			No closed-form solution to the maximization problem is available. So, the maximum likelihood  estimator of $\boldsymbol{\theta}_\rho$ can only be obtained via numerical optimization.

		}

		%\newpage
		\subsection{Maximum product of spacings method}\label{Method of maximum product of spacings}
		
		The maximum product of spacings (MPS) method was introduced by \cite{ChengAmin1979,ChengAmin1983} and \cite{Ranneby1984} as a different approach to estimating parameters of continuous univariate distributions compared to maximum likelihood estimation. Let $\{Z_i:i = 1,\ldots,n\}$ be a univariate random sample of size $n$ from the {\color{black} UBBS1}$(\boldsymbol{\theta}_\rho)$
		distribution with PDF as in \eqref{main-pdf}, and let $z_i$ be the sample observation of $Z_i$. Then, the uniform spacings of  $\{z_i:i = 1,\ldots,n\}$ are given by
		\begin{eqnarray*}
			\Delta_{i}(\boldsymbol{\theta}_\rho)=F_Z(z_{i:n};\boldsymbol{\theta}_\rho)-
			F_Z(z_{i-1:n};\boldsymbol{\theta}_\rho), \quad  1,\ldots,n,
		\end{eqnarray*}
		with $F_Z(z_{0:n};\boldsymbol{\theta}_\rho)=0$, $F_Z(z_{n+1:n};\boldsymbol{\theta}_\rho)=1$,
		and $\Delta_{0}(\boldsymbol{\theta}_\rho)+\Delta_{1}(\boldsymbol{\theta}_\rho)+\cdots+\Delta_{n+1}(\boldsymbol{\theta}_\rho)=1$.
		
		The maximum product of spacings estimator, $\widehat{\boldsymbol{\theta}}_\rho$, is obtained by maximizing the geometric mean of the spacings
		\begin{eqnarray*}
			G(\boldsymbol{\theta}_\rho)=\left( \prod_{i=1}^{n+1} \Delta_{i}(\boldsymbol{\theta}_\rho) \right)^{\frac{1}{n+1}},
		\end{eqnarray*}
		or, equivalently, maximizing the function
		\begin{eqnarray*}
			H(\boldsymbol{\theta}_\rho)= \frac{1}{1+n} \sum_{i=1}^{n+1}\log( \Delta_{i}(\boldsymbol{\theta}_\rho)).
		\end{eqnarray*}

		%the log-likelihood function for $\boldsymbol{\theta}_\rho=(\alpha_1,\alpha_2,\beta_1,\beta_2,\rho)^\top$, $%\alpha_1> 0,\alpha_2> 0 ,\beta_1> 0, \beta_2> 0$ and $\rho\in(-1,1)$, is given by (without additive constant)

		\section{Simulation study}\label{sec:04}

		In this section, we carry out a Monte Carlo simulation study for evaluating the performance of the above described {\color{black}maximum likelihood and maximum product of spacings methods} for the {\color{black} UBBS1} distribution \eqref{main-pdf}. The simulation scenario considers the following setup: 300 Monte Carlo replications, sample size $n \in \{100,200,400,800\}$, and vector of true parameters $(\alpha_1,\alpha_2,\beta_1,\beta_2)^\top= (0.5,0,5,1.0,1.0)^\top$, $\rho \in \{0.10,0.25,0.5,0.75\}$. {\color{black}A generator of random numbers for the UBBS1 distribution is summarized in Algorithm \ref{alg:rn}.}

		\begin{algorithm}[h!]
			\caption{\color{black}Generator of random numbers from a UBBS1 distribution}\label{alg:rn}
			\begin{algorithmic}[1]
				\State Generate $n$ random numbers from the bivariate normal distribution, $\boldsymbol{X}=(X_1,X_2)^{\top}\sim\textrm{N}_2\left(\boldsymbol{\mu}=(0,0)^{\top},\boldsymbol{\Sigma}=\left(
				\begin{matrix}
					1 & \rho  \\
					\rho & 1 \\
				\end{matrix}
				\right)\right)$, as
				\begin{itemize}
					\item Generate a matrix $\boldsymbol{W}$ of size $n \times 2$ containing $2n$ random numbers from $\text{N}(0,1)$.
					
					\item
					Compute the decomposition $\boldsymbol{\Sigma} = \boldsymbol{Q}^{\top} \boldsymbol{Q}$.
					
					\item
					Apply the transformation $\boldsymbol{X} = \boldsymbol{W} \boldsymbol{Q} + \boldsymbol{J} \boldsymbol{\mu}^{\top}$, where $\boldsymbol{J}$ is a column vector of ones.
					
					\item
					Obtain the matrix $\boldsymbol{X}=(X_1,X_2)^{\top}$ of size $n \times 2$.
					
				\end{itemize}
				
				\State
				Compute $n$ random numbers from $T_i \sim \textrm{BS}(\alpha_i,\beta_i)$ by using
				$$
				T_i = \beta_i\left(\frac{\alpha_i X_i}{2} + \sqrt{\frac{\alpha_i^2X_i^2}{4} +1}\right)^2, \quad i=1,2.
				$$
				
				\State
				Compute $n$ random numbers from $Z\sim {\rm {\color{black} UBBS1}}(\boldsymbol{\theta}_\rho=(\alpha_1,\alpha_2,\beta_1,\beta_2,\rho)^{\top})$ by using
				$Z=\frac{T_1}{T_1+T_2}$.
				
			\end{algorithmic}
		\end{algorithm}

		{\color{black}The performance and recovery of the estimators} are evaluated by means of relative bias (RB), and root mean square error (RMSE), which are calculated from the Monte Carlo replicates, as
		\begin{eqnarray*}
			\widehat{{\color{black}\textrm{RB}}}(\widehat{\gamma}) &=&  \frac{1}{N} \sum_{i = 1}^{N}\left| \frac{(\widehat{\gamma}^{(i)} - \gamma)}{\gamma}\right| ,\quad
			\widehat{\mathrm{RMSE}}(\widehat{\gamma}) = {\sqrt{\frac{1}{N} \sum_{i = 1}^{N} (\widehat{\gamma}^{(i)} - \gamma)^2}}, \\
		\end{eqnarray*}
		where $\gamma$ and $\widehat{\gamma}^{(i)}$ are the true parameter value and its $i$-th estimate, and $N$ is the number of Monte Carlo replications. The steps for the Monte Carlo simulation study {\color{black} are as follows}:
		i. set the values of the parameters of the {\color{black} UBBS1} distribution; ii. generate 300 samples of size $n$ from the chosen model; iii. estimate the model parameters using the maximum product of spacings method for each sample; and iv. compute the empirical RB and RMSE.
		
\begin{figure}[!ht]
			\vspace{-0.25cm}
			\centering
			{\includegraphics[height=4.0cm,width=4.0cm]{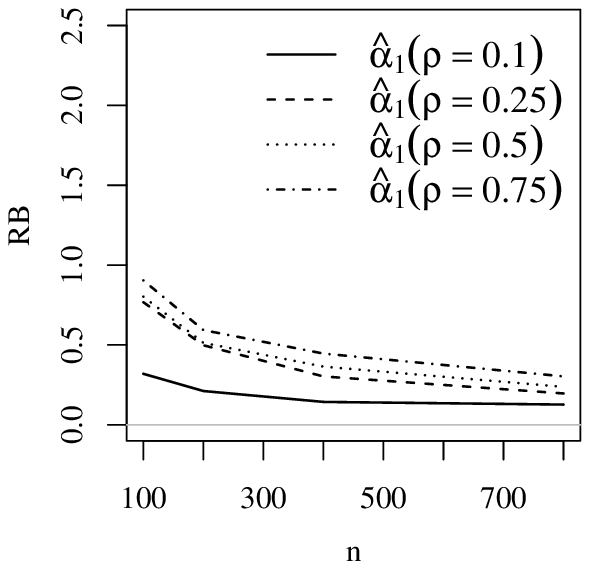}}\hspace{-0.10cm}
			{\includegraphics[height=4.0cm,width=4.0cm]{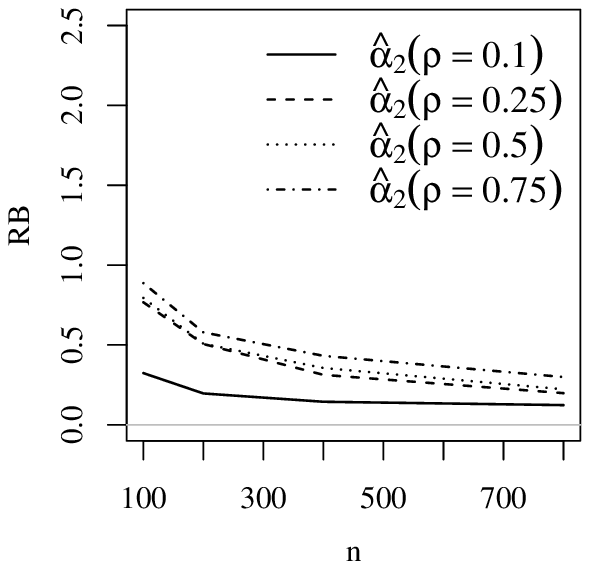}}\hspace{-0.10cm}
			{\includegraphics[height=4.0cm,width=4.0cm]{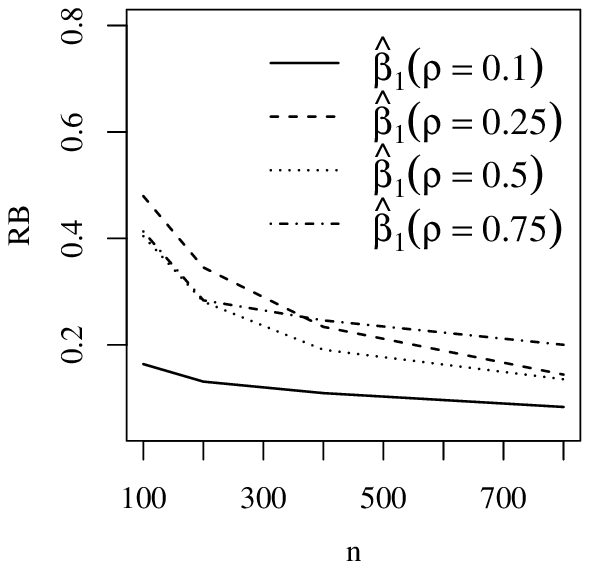}}\hspace{-0.10cm}
			{\includegraphics[height=4.0cm,width=4.0cm]{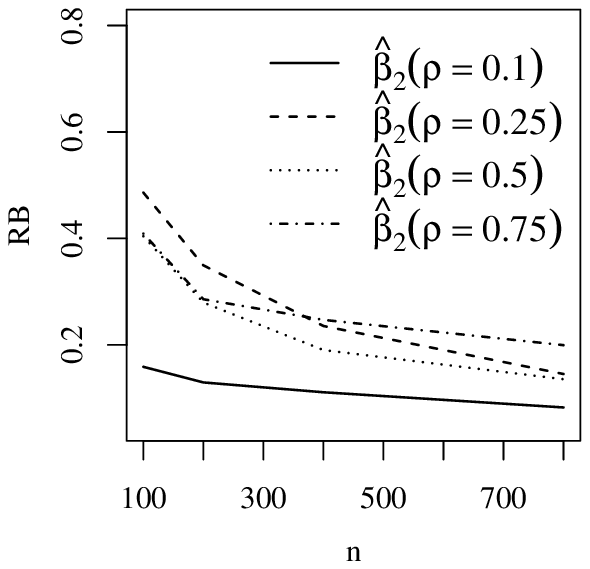}}\hspace{-0.10cm}
			{\includegraphics[height=4.0cm,width=4.0cm]{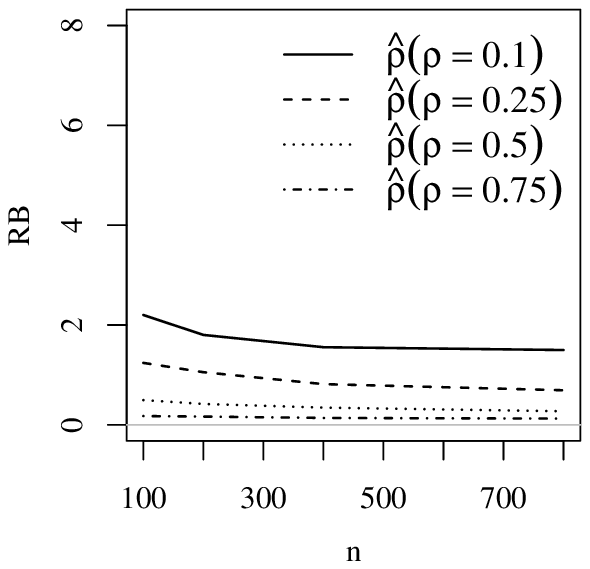}}
			\vspace{-0.1cm}
			\caption{{\color{black}Monte Carlo simulation results for the  UBBS1 distribution based on the maximum product of spacings method.}}
			\label{fig_normal_mc1}
		\end{figure}

\begin{figure}[!ht]
			\vspace{-0.25cm}
			\centering
			{\includegraphics[height=4.0cm,width=4.0cm]{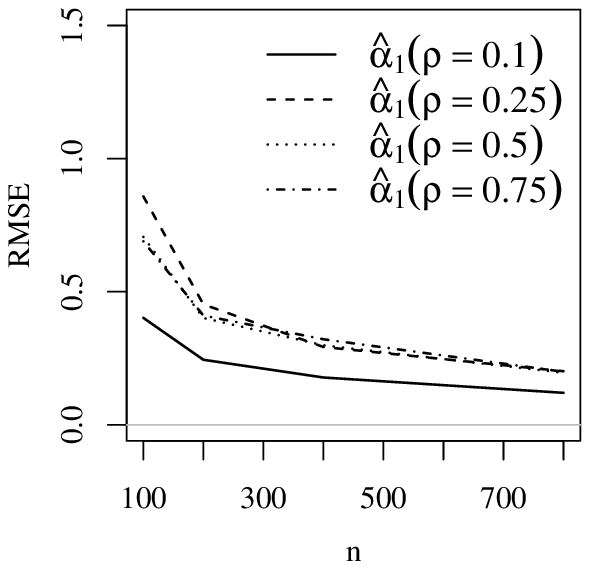}}\hspace{-0.10cm}
			{\includegraphics[height=4.0cm,width=4.0cm]{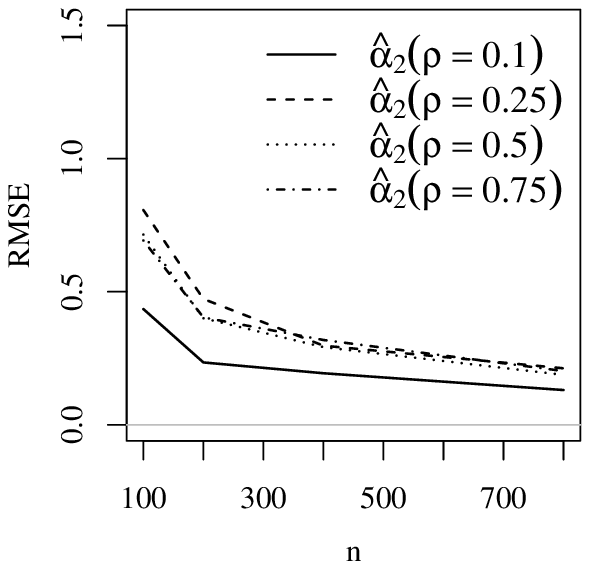}}\hspace{-0.10cm}
			{\includegraphics[height=4.0cm,width=4.0cm]{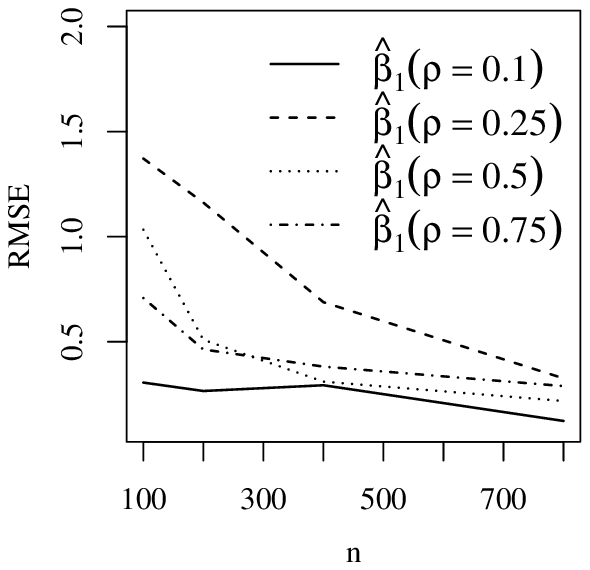}}\hspace{-0.10cm}
			{\includegraphics[height=4.0cm,width=4.0cm]{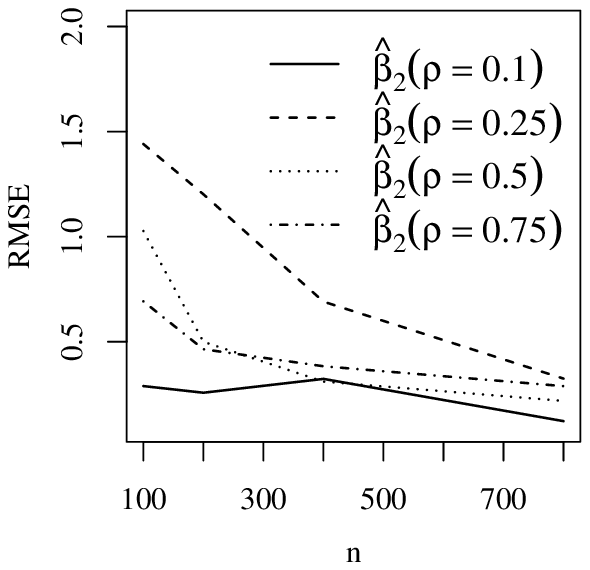}}\hspace{-0.10cm}
			{\includegraphics[height=4.0cm,width=4.0cm]{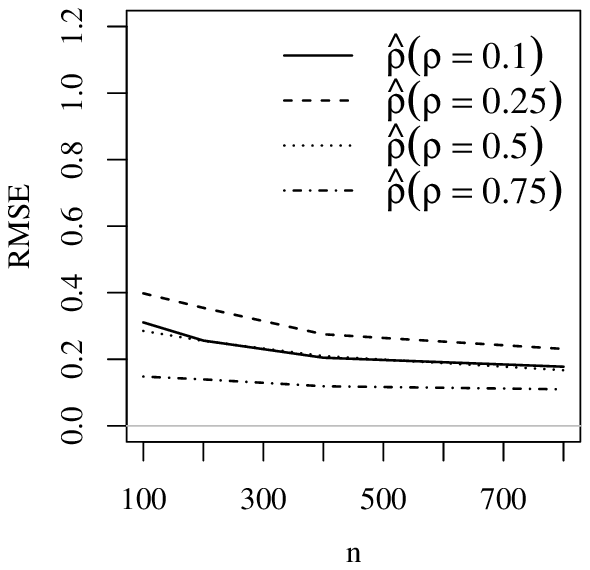}}
			\vspace{-0.1cm}
			\caption{{\color{black}Monte Carlo simulation results for the  UBBS1 distribution based on the maximum product of spacings method.}}
			\label{fig_normal_mc2}
		\end{figure}

\begin{figure}[!ht]
			\vspace{-0.25cm}
			\centering
			{\includegraphics[height=4.0cm,width=4.0cm]{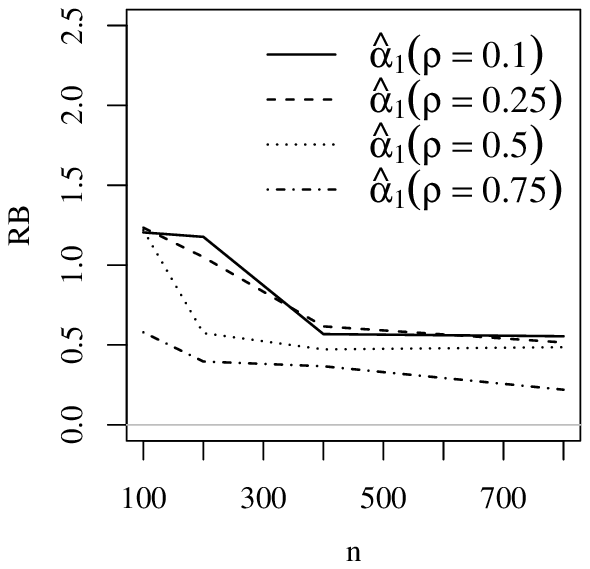}}\hspace{-0.10cm}
			{\includegraphics[height=4.0cm,width=4.0cm]{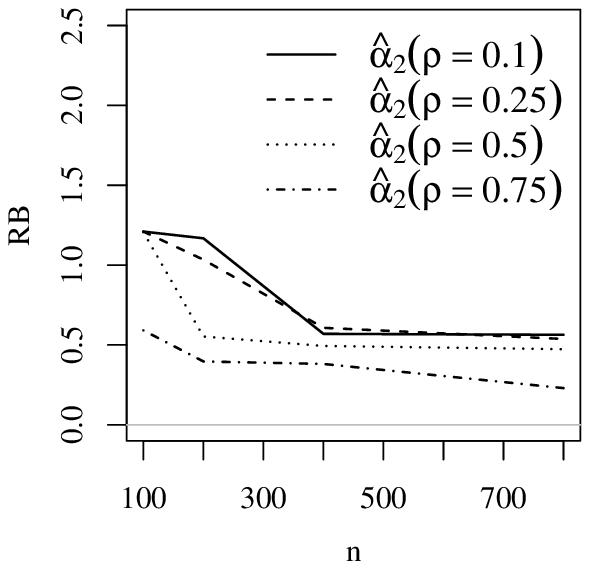}}\hspace{-0.10cm}
			{\includegraphics[height=4.0cm,width=4.0cm]{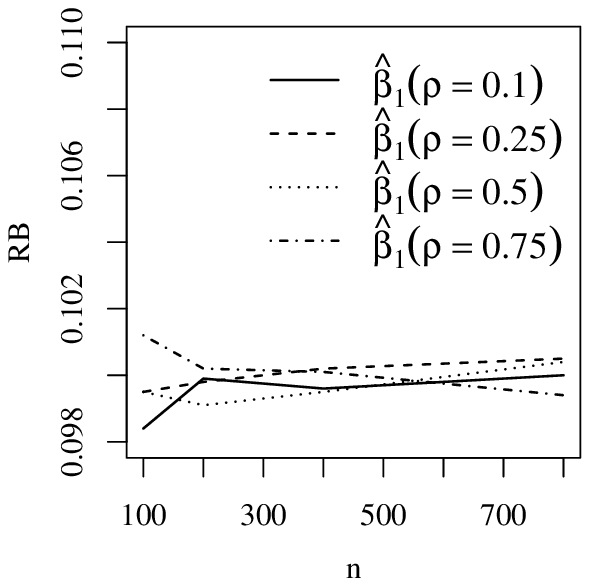}}\hspace{-0.10cm}
			{\includegraphics[height=4.0cm,width=4.0cm]{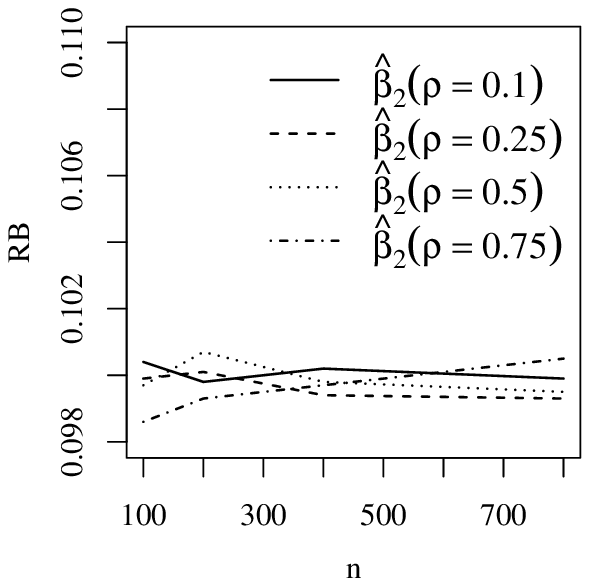}}\hspace{-0.10cm}
			{\includegraphics[height=4.0cm,width=4.0cm]{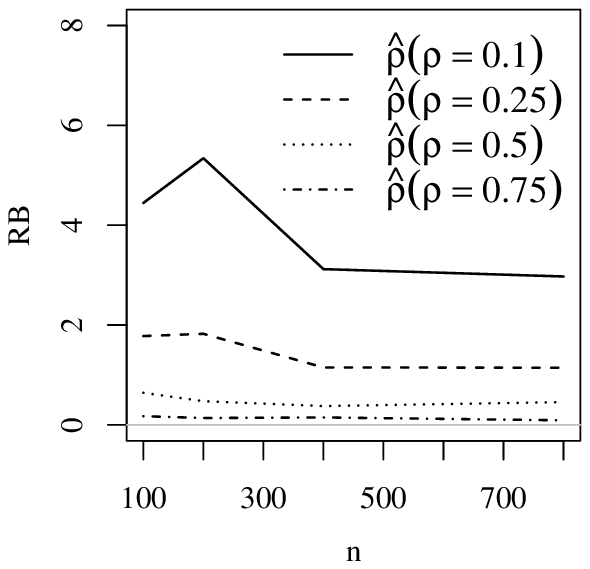}}
			\vspace{-0.1cm}
			\caption{{\color{black}Monte Carlo simulation results for the  UBBS1 distribution based on the maximum likelihood method.}}
			\label{fig_normal_mc3}
		\end{figure}

\begin{figure}[!ht]
			\vspace{-0.25cm}
			\centering
			{\includegraphics[height=4.0cm,width=4.0cm]{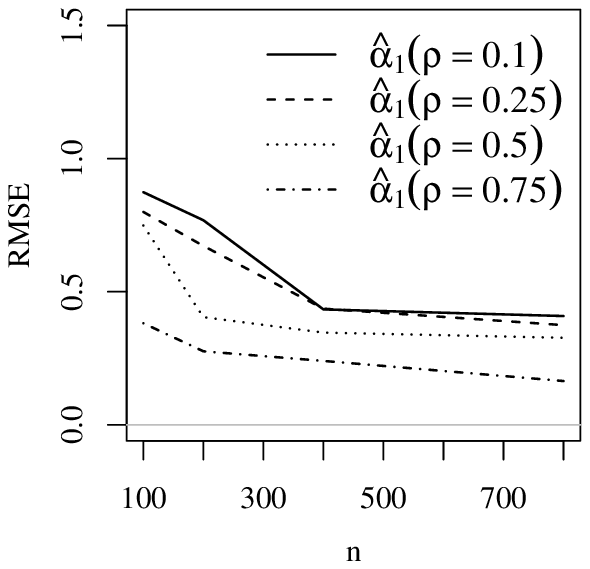}}\hspace{-0.10cm}
			{\includegraphics[height=4.0cm,width=4.0cm]{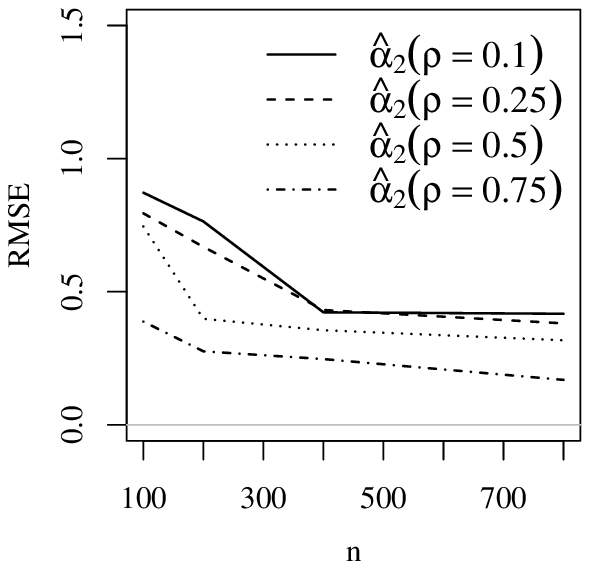}}\hspace{-0.10cm}
			{\includegraphics[height=4.0cm,width=4.0cm]{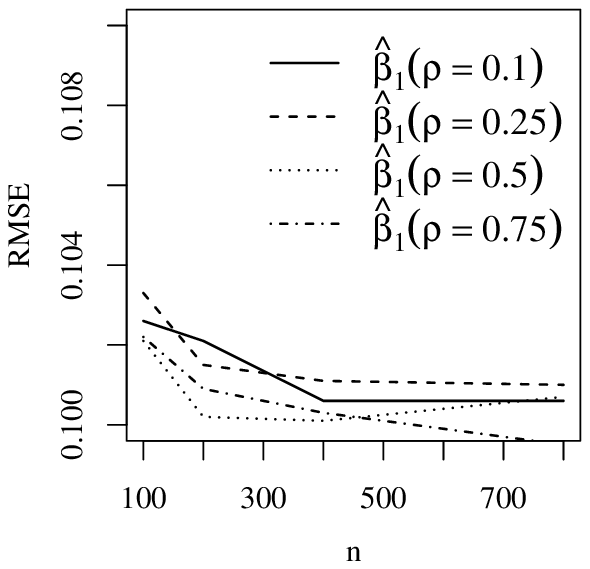}}\hspace{-0.10cm}
			{\includegraphics[height=4.0cm,width=4.0cm]{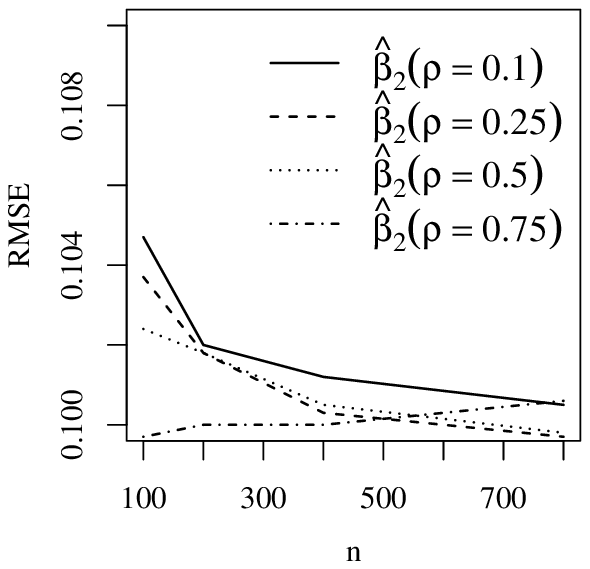}}\hspace{-0.10cm}
			{\includegraphics[height=4.0cm,width=4.0cm]{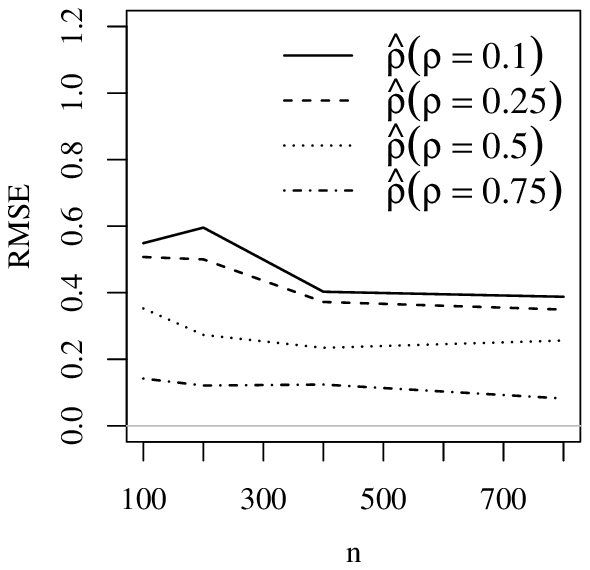}}
			\vspace{-0.1cm}
			\caption{{\color{black}Monte Carlo simulation results for the  UBBS1 distribution based on the maximum likelihood method.}}
			\label{fig_normal_mc4}
		\end{figure}

		{\color{black}The maximum likelihood and maximum product of spacings estimation results so obtained are shown in Figures \ref{fig_normal_mc1}--\ref{fig_normal_mc4}, wherein the empirical RB and RMSE are both plotted. The following results are observed: i) In general, when the sample size increases, both RB and RMSE decrease for all estimators and cases; ii) in the maximum product of spacings case, we observe that the RB of $\widehat{\alpha}_1$, $\widehat{\alpha}_2$, $\widehat{\beta}_1$, and $\widehat{\beta}_2$ tends to increase as $\rho$ increases. In the case of $\widehat{\rho}$, we observe the opposite behavior; and iii) in the maximum likelihood case, the impact of $\rho$ on the RB of $\widehat{\alpha}_1$, $\widehat{\alpha}_2$, $\widehat{\beta}_1$, $\widehat{\beta}_2$ are somewhat undefined; iv) the RB for $\widehat{\alpha}_1$, $\widehat{\alpha}_2$, and $\widehat{\rho}$ is smaller for the estimators based on the maximum product of spacings method compared to maximum likelihood, while for $\widehat{\beta}_1$ and $\widehat{\beta}_2$, the RB is smaller for the estimators based on the maximum likelihood method. We observe a similar behavior in the RMSE case, except for $\widehat{\rho}$, where the maximum product of spacings method performs better.}

		{\color{black}We now compare the performance of the maximum likelihood and maximum product of spacings methods when bimodality is present. We consider $\bm\theta_\rho = (1.6, 0.7, 1.1, 0.9, 0.6)^\top$ with $n=100$, that is, the density exhibits a bimodal behavior; see Table \ref{tab_sim_comp1}. The results show that, with the exception of the $\widehat{\beta}_2$ case, the performance is better for estimates based on the MPS method; we considered other sample sizes and found similar results.}

		% latex table generated in R 4.4.1 by xtable 1.8-4 package
		% Sat Oct 26 12:31:35 2024
		\begin{table}[ht]
			\centering
			\caption{
				{\color{black}RB and RMSE (between brackets) from simulated data for the UBBS1 parameter estimates based on the maximum likelihood (ML) and maximum product of spacings (MPS) methods.}}
			\begin{tabular}{lllllllllll}
				\hline
				Method &  & $\widehat{\alpha}_1$ & $\widehat{\alpha}_2$ & $\widehat{\beta}_1$ & $\widehat{\beta}_2$  & $\widehat{\rho}$ \\
				\hline
				MPS     &     & 0.1523   & 0.1608  & 0.0767  & 0.3426  & 0.3074\\
				&      & (0.2437) &(0.1125) & (0.0843)& (0.3084)& (0.1844)\\
				ML      &     & 0.2721   & 0.8957  & 0.1714  & 0.0111  & 1.6970 \\
				&      & (0.4353) & (0.6270)& (0.1885)& (0.0100)&(1.0182) \\
				
				\hline
			\end{tabular}\label{tab_sim_comp1}
		\end{table}

		%\newpage

		\section{Application}\label{sec:05}

		We now give two applications using real data previously analyzed in the literature. We study the validity of the new model for the data sets and show that {\color{black} UBBS1} distribution presents good fits in both cases.

		\subsection{Income-consumption data}

		In order to evaluate the new probabilistic model, a public data set of income and consumption data was used.
		The data (available at \url{https://www.bancaditalia.it/statistiche/tematiche/indagini-famiglie-imprese/bilanci-famiglie/documentazione/ricerca/ricerca.html?min_anno_pubblicazione=2008&max_anno_pubblicazione=2008}, accessed on July 22, 2024) comes from the Bank of Italy's Survey on Household Income and Wealth from the year 2008.

		The income comprises payroll income, pensions and net transfers, net self-employment income, and property income. 
		% \begin{itemize}
			%     \item payroll income (net wages, salaries and fringe benefits);
			%     \item pensions and net transfers (pensions, arrears and other transfers);
			%     \item net self-employment income (self-employment income and entrepreneurial income); and
			%     \item property income (either from real estate or from financial assets).
			% \end{itemize}
		Consumption consists of expenditures on both durable and non-durable goods. Durable expenditure represents the net balance of purchased and sold goods. Non-durable expenditures include monthly expenses such as rent, food, and other essentials, as well as annual expenses.
		
		We used the data from the survey, available as the data sets RFAM08 (income as $Y$ variable) and RISFAM08 (expenditures as $C$ variable), according to the Survey on Household Income and Wealth 2008 data description file.
		We removed from the data set lines whose income or consumption was negative or unavailable.
		Thus, the analyzed data set included information from 7,957 families, after removing 20 lines of the data.
		
		The same data set was also used in \cite{domma2012stress} and \cite{Lima2024assessing}, however, they used a copula approach to model the data and compared income and consumption through stress-strength measures of the type $\mathbb{P}(X<Y)$.
		Our framework involves constructing a new variable $U=X/(X+Y)$.
		In summary, when $X$ and $Y$ represent expenditures and income, respectively,  $U < 1/2$ indicates that the family ended the year with more expenditures than income. If $U > 1/2$, the opposite is true. The case $U = 1/2$ means expenditures and income are equal.

\begin{figure}[!ht]
			\centering
			\includegraphics[width=1.0\linewidth]{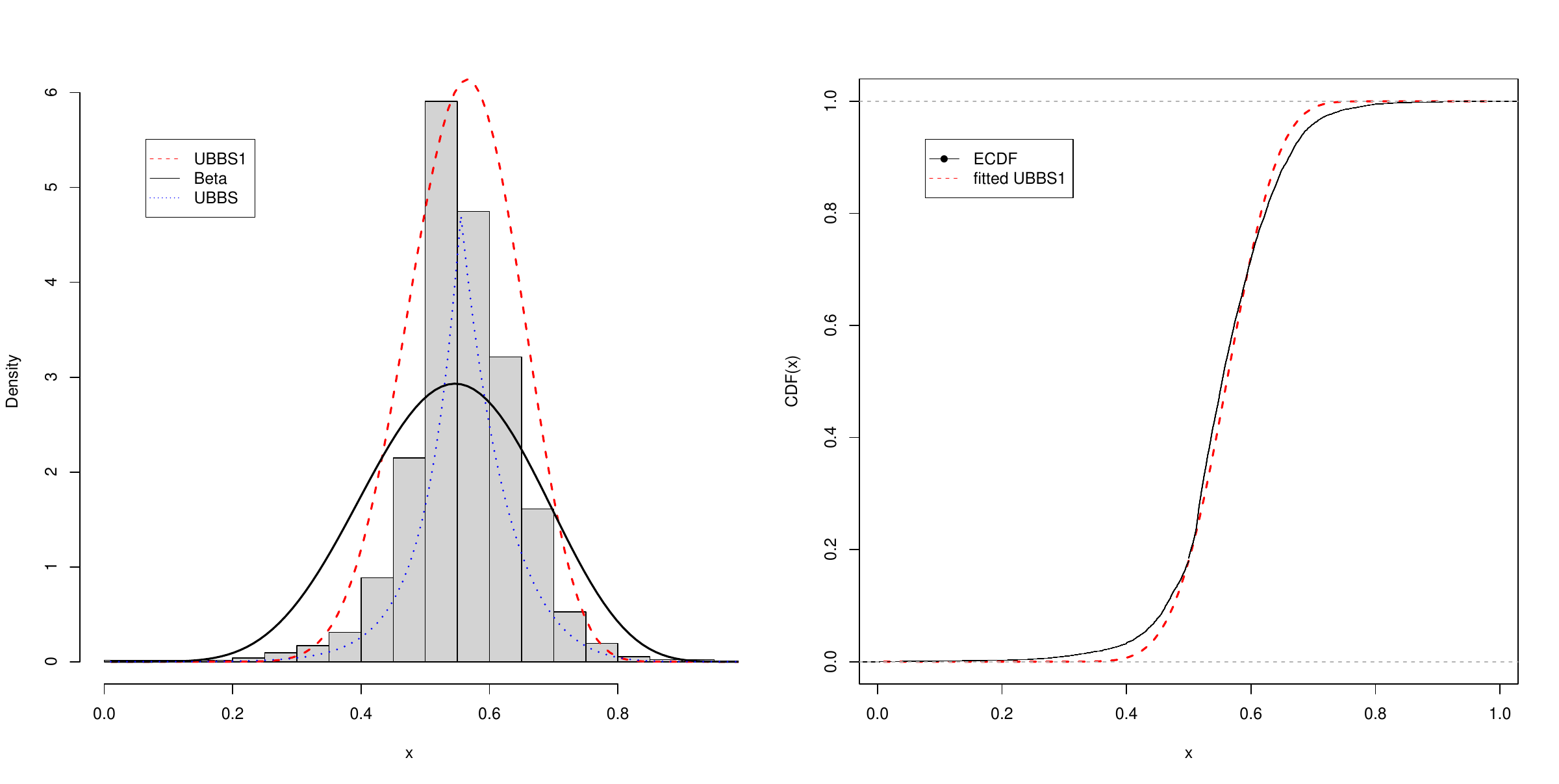}
			\caption{Fitted {\color{black} UBBS1} PDFs (left) and empirical CDF (right) for income-consumption data.}
			\label{fig:ic_hist_ecdf}
		\end{figure}

		Table \ref{tab:descritiva_receitas_despesas} presents descriptive statistics for $U=X/(X+Y)$.
		The estimated parameters of {\color{black} UBBS1} model was $\widehat{\bm \theta}_\rho = (0.275, 0.274, 1.041, 1.331, 0.149)$.
		Figure \ref{fig:ic_hist_ecdf} shows the fit of the {\color{black} UBBS1} distribution to the density and ECDF of the data. The histogram of $U$ shows that with high (81.5\%) probability $U>1/2$. That means, in general, the families have income greater than expenditures.

		% latex table generated in R 4.2.2 by xtable 1.8-4 package
		% Mon May  1 13:24:37 2023
		\begin{table}[!ht]
			\centering
			\caption{Descriptive statistics for income-consumption data.}
			\label{tab:descritiva_receitas_despesas}
			%\resizebox{.95\linewidth}{!}{
				\begin{tabular}{crrrrrrrrr}
					\hline
					n & Min. & 1st Qu. & Median & Mean & 3rd Qu. & Max. & Std. dv. & CS & CK\\ 
					\hline
					7,957 &  0.004
					&  0.514
					& 0.553
					&  0.557
					&  0.608
					& 0.935
					& 0.086
					& -0.419
					& 3.343\\
					\hline
				\end{tabular}
				%}
		\end{table}

		% \begin{figure}[H]
			%     \centering
			%     \includegraphics[width=1.0\linewidth]{figs/re_qqnorm_residuals.pdf}
			%     \caption{Normal Q-Q and P-P plots for the residuals.}
			%     \label{fig:ic_qqplots}
			% \end{figure}
		
		%\textcolor{magenta}{Analyzing Figure 12, we conclude that, in general, the residuals are well-behaved.}
		
		Given that the $U$ data has unit support, we can make a comparison between the utilization of the {\color{black} UBBS1} model and other families of unit distributions, such as Beta and the unit bimodal BS (UBBS\footnote{The UBBS model was proposed in \cite{Martinez2024unit}.}) models, as illustrated in Table \ref{tab:ic_AIC_BIC}.
		To obtain the estimation results for Beta and UBBS models in Table \ref{tab:ic_AIC_BIC}, we used the package \textit{AdequacyModel} of the R software \textcolor{black}{ and we used the function '\textit{goodness.fit}' to obtain maximum likelihood estimates}.
		Deciding the better distribution to model the data was oriented by AIC (Akaike information criterion) and BIC (Bayesian information criterion) criteria.
		In Table \ref{tab:ic_AIC_BIC}, these criteria suggest {\color{black} UBBS1} fits the data better than Beta and UBBS models.
		
		\begin{table}[!ht]
			\caption{Estimated parameters for income-consumption data and model selection with AIC and BIC.}
			\label{tab:ic_AIC_BIC}
			\centering
			\begin{tabular}{ccccc}
				\hline
				Model & $\hat{\bm\theta}$ & $ll_{max}$ & AIC & BIC  \\ 
				\hline
				{\color{black} UBBS1} & (0.275, 0.274, 1.041, 1.331, 0.149) & -7,062.74 & -14,135.49 &  -14,170.40 \\ 
				Beta & (7.492, 6.390) &  -6,877.40 & -13,750.79 & -13,736.83 \\ 
				UBBS & (1.726, 0.589, 9.545) &  -3,614.60 & -7,223.20 & -7,202.25 \\ 
				\hline
			\end{tabular}
		\end{table}

		\subsection{Body mass data}

\begin{figure}[!ht]
			\centering
			\includegraphics[width=1.0\linewidth]{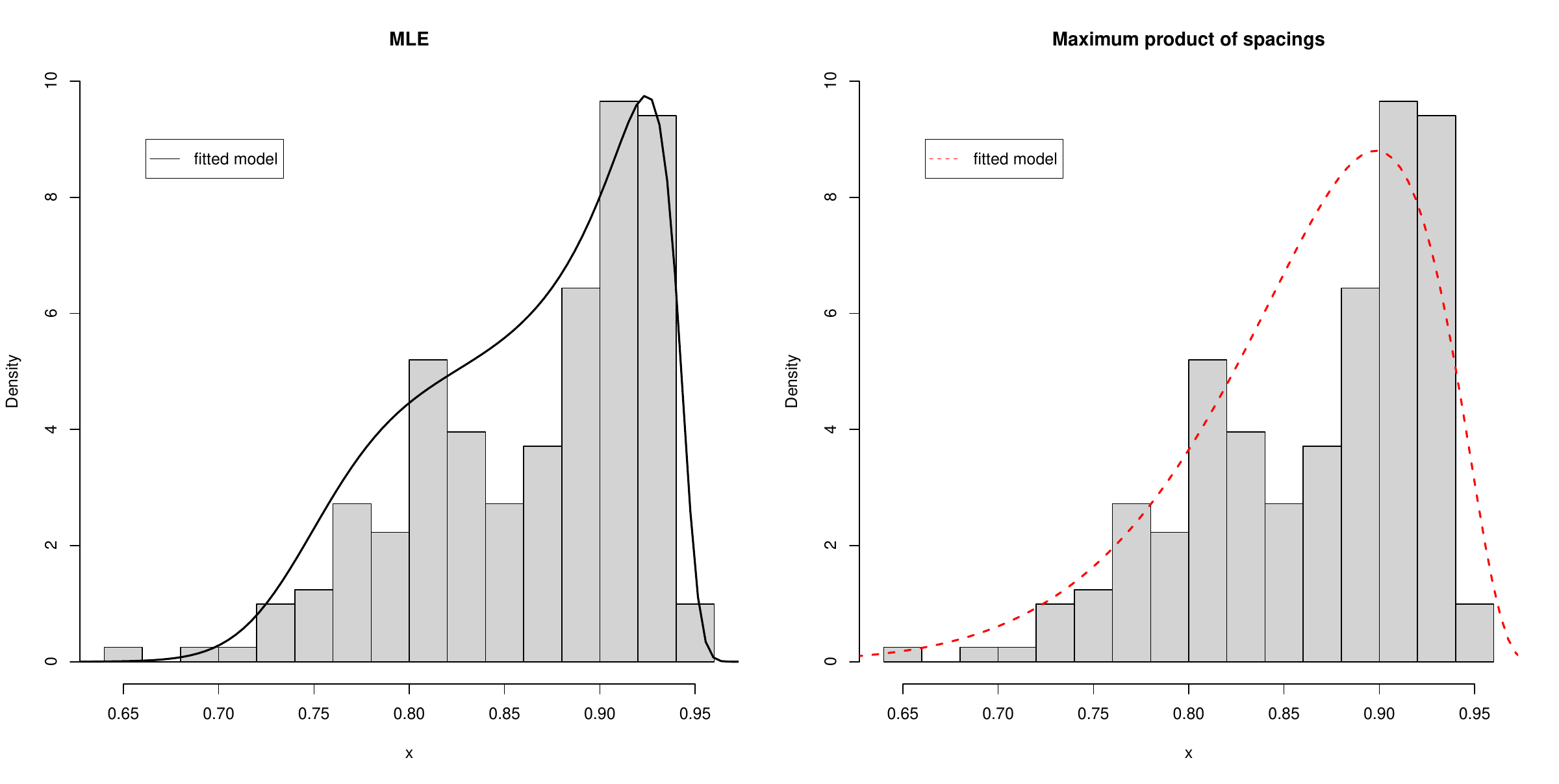}
			\includegraphics[width=1.0\linewidth]{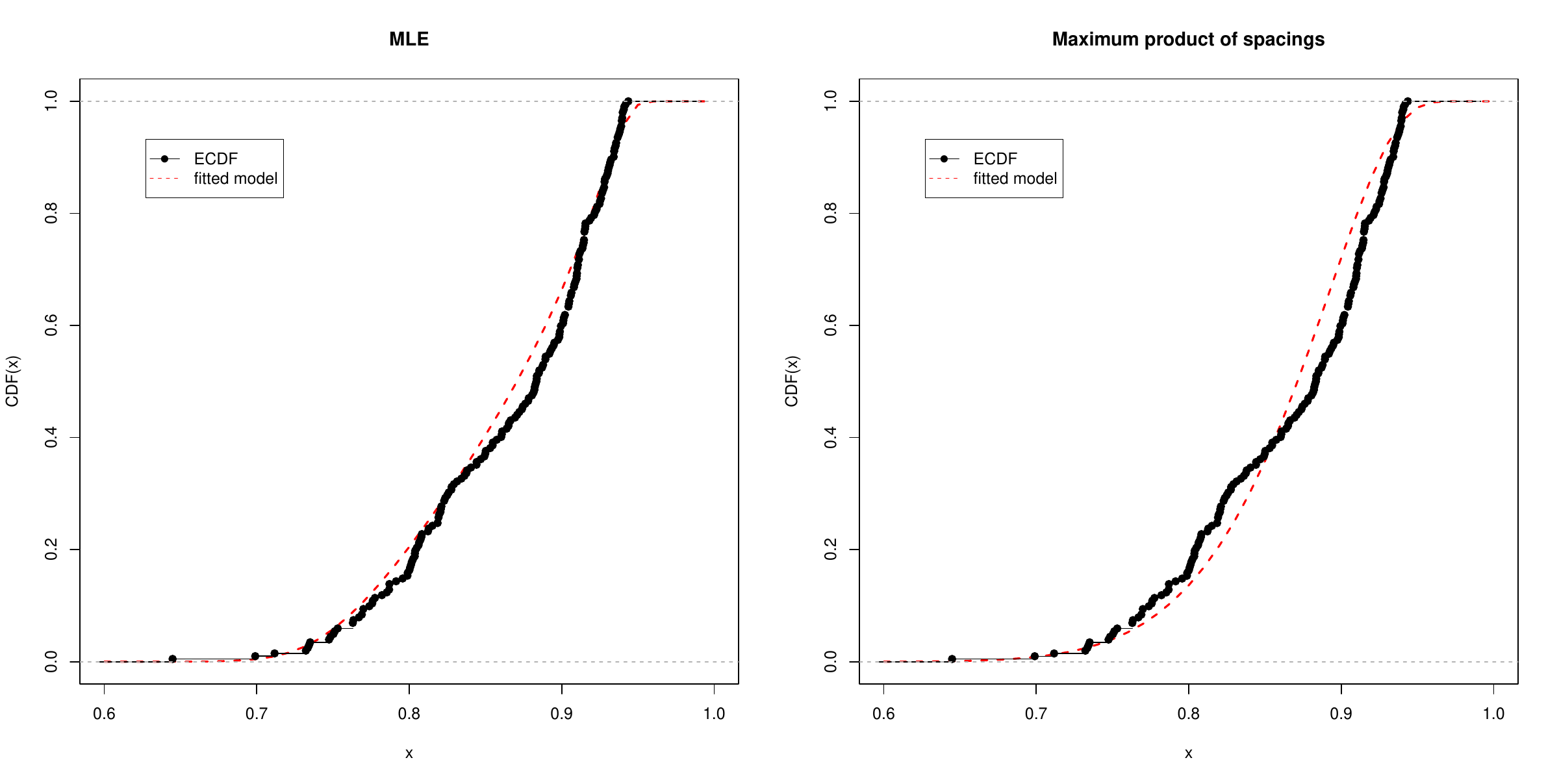}
			\caption{Plots for the athlete's body mass. On top, histogram and fitted {\color{black} UBBS1} PDFs (ML estimates on left and Maximum product of spacings estimates on right); on bottom, Empirical CDF and fitted {\color{black} UBBS1} CDF.} 
			\label{fig:body_hist_ecdf}
		\end{figure}

		Biomedical measurements of 202 athletes from different sports are presented in the Australian Institute of Sport (AIS) data set. The authors in \cite{Martinez2024unit} proposed the UBBS model and showed its fit on the distribution of the percentage of the body fat complement of the athletes. In this subsection, we show that the new {\color{black} UBBS1} model is an alternative distribution to fit that data.
		The data were imported directly through the software R by the command:
		\begin{verbatim}
			library(sn)
			data(ais)
			u=1-(ais$Bfat/100)
		\end{verbatim}
		
		Descriptive statistics for $\mathbf{u}$ (the athlete's body mass in the data set) are presented in Table \ref{tab:body_summary}.
		As the data set has unit support, the {\color{black} UBBS1} distribution is a candidate to model such data. 
		Figure \ref{fig:body_hist_ecdf} shows the fit of {\color{black} UBBS1} to the data.
		Maximum likelihood estimates and maximum product of spacings estimates can be compared graphically. 
		\textcolor{black}{To obtain the ML estimates, we minimized the minus log-likelihood function using the \textit{optim} function in the R software.}
		
		\begin{table}[!ht]
			\caption{Descriptive statistics of athlete's body mass in the data set.}
			\centering
			\begin{tabular}{rrrrrrrrr}
				\hline
				Min. & 1st Qu. & Median & Mean & 3rd Qu. & Max. & Std. dv. & CS & CK  \\ 
				\hline
				0.645
				& 0.819
				& 0.884
				& 0.865
				& 0.915
				& 0.944
				& 0.062
				&-0.754
				&-0.201
				\\ 
				\hline
			\end{tabular}
			\label{tab:body_summary}
		\end{table}

		In Table \ref{tab:body_AIC_BIC}, when comparing the {\color{black} UBBS1} model (with parameter estimates via product of spacings) with the Beta, UBBS, \textcolor{black}{and Bimodal Beta (BBeta)} models using information criteria, the BIC indicates the new model is better to fit the athlete's body mass data. However, when considering the AIC, the UBBS model should be chosen.
		\textcolor{black}{
			The ML estimates for the Beta and UBBS models, presented in Table \ref{tab:body_AIC_BIC}, were previously reported in Table 6 of \cite{Martinez2024unit}. For the BBeta model \cite{Vila2024a}, the ML estimates were obtained using the '\textit{optim}' function.
		}

		\begin{table}[!ht]
			\caption{Estimated parameters for the athlete's body mass data and model selection with AIC and BIC.}
			\label{tab:body_AIC_BIC}
			\centering
			\begin{tabular}{cccc}
				\hline
				Model & $\hat{\bm\theta}$ & AIC & BIC  \\ 
				\hline
				\textcolor{black}{UBBS1 (MPS)} & (0.149, 0.626, 0.296, 2.003, 0.586)  & -613.10 &  -629.64 \\ 
				\textcolor{black}{UBBS1 (MLE)} & \textcolor{black}{(1.763, 1.106, 0.163, 1.097, 0.981)} & \textcolor{black}{-619.30} & \textcolor{black}{-602.75} \\
				Beta & (0.865, 0.173) &   -585.32 & -578.70 \\ 
				UBBS & (0.284, 0.137, -1.353) &  -626.64 & -616.72 \\ 
				\textcolor{black}{BBeta} & \textcolor{black}{(26.1007122   4.3747393   0.9712073 -17.9150416)} & \textcolor{black}{-581.3334} & \textcolor{black}{-568.1003} \\
				\hline
			\end{tabular}
		\end{table}
		
		\section{Concluding remarks}\label{sec:06}
		
		In this paper, we have introduced a new unit distribution. Its generation as a ratio of possibly dependent Birnbaum-Saunders random variables has been discussed in detail. As a more theoretical part, we have derived the cumulative distribution function {\color{black} to then obtain simple expressions of the moment-generating function, moments } and the stress-strength probability. As practical issues, we presented maximum product of spacings estimators and applications to income-consumption and body mass data sets.
		
		The research could be continued in several directions. One could investigate {\color{black} the entropy of the distribution and use the moments formulas in \eqref{moments}
			%
			%the moments and the moment generating function 
			to} try to express shape characteristics like skewness and kurtosis. 
		%The analysis of the entropy of the distribution could also be of interest. 
		Instead of using the ratio $Y/(X+Y)$ one could explore other construction principles like those mentioned in the introduction to obtain a unit distribution. Finally, one could use a bivariate distribution other than the Birnbaum-Saunders to construct a unit distribution.

		%%%%%%%%%%%%%%%%%%%%%%%%%%%%%%%%%%%%%%%%%%%%%%%%%%%%%%%%%%%%%

	%\newpage 	
	\paragraph{Acknowledgements}
	This study was financed in part by the
	Coordenação de Aperfeiçoamento de Pessoal de Nível Superior - Brasil (CAPES) - Finance Code 001.
	
	\paragraph{Disclosure statement}
	There are no conflicts of interest to disclose.

	%%%%%%%%%%%%%%%%%%%%%%%%%%%%%%%%%%%%%%%%%%%%%%%%%%%%%%%%%%%%%

%	\bibliographystyle{unsrt}
	%	\bibliography{BibTexFileName}

	%\typeout{}
	%\bibliography{ref}

	%\newpage
	%\begin{appendices}
	%
	%\end{appendices}
	
\end{document}